\documentclass[]{iopart}

\usepackage{xcolor}
\usepackage{graphicx}
\usepackage{dcolumn}
\usepackage{bm}
\usepackage{booktabs}    
\usepackage{multirow}     
\usepackage{subfigure}
\usepackage[para,online,flushleft]{threeparttable}
\usepackage{textcomp}
\usepackage[utf8]{inputenc}
\usepackage[T1]{fontenc}
\usepackage{mathptmx}
\usepackage{rotating}
\usepackage{pdflscape,caption}

\usepackage{iopams}
\usepackage{cite}

\usepackage{geometry}
\begin{document}

\title{A Combined Density Functional Theory and X-ray Photoelectron Spectroscopy Study of the Aromatic Amino Acids}

\author{Anna Regoutz}
 \ead{a.regoutz@ucl.ac.uk}
\address{ 
Department of Chemistry, University College London, 20 Gordon Street, London, WC1H~0AJ, United Kingdom.
}
\author{Marta S. Wolinska}
\address{ 
Department of Materials, Imperial College London, London SW7 2AZ, United Kingdom.
}
\author{Nathalie K. Fernando}
\address{ 
Department of Chemistry, University College London, 20 Gordon Street, London, WC1H~0AJ, United Kingdom.
}
\author{Laura E. Ratcliff}
 \ead{laura.ratcliff08@imperial.ac.uk}
\address{ 
Department of Materials, Imperial College London, London SW7 2AZ, United Kingdom.
}

\date{\today}

\begin{abstract}
Amino acids are essential to all life. However, our understanding of some aspects of their intrinsic structure, molecular chemistry, and electronic structure is still limited. In particular the nature of amino acids in their crystalline form, often essential to biological and medical processes, faces a lack of knowledge both from experimental and theoretical approaches. An important experimental technique that has provided a multitude of crucial insights into the chemistry and electronic structure of materials is X-ray photoelectron spectroscopy. Whilst the interpretation of spectra of simple bulk inorganic materials is often routine, interpreting core level spectra of complex molecular systems is complicated to impossible without the help of theory. We have previously demonstrated the ability of density functional theory to calculate binding energies of simple amino acids, using $\Delta$SCF implemented in a systematic basis set for both gas phase (multiwavelets) and solid state (plane waves) calculations. In this study, we use the same approach to successfully predict and rationalise the experimental core level spectra of phenylalanine (Phe), tyrosine (Tyr), tryptophan (Trp), and histidine (His) and gain an in-depth understanding of their chemistry and electronic structure within the broader context of more than 20 related molecular systems. The insights gained from this study provide significant information on the nature of the aromatic amino acids and their conjugated side chains.
\end{abstract}

\section{\label{sec:intro}Introduction}

\begin{figure}[h]
\centering
\includegraphics[width=0.6\textwidth]{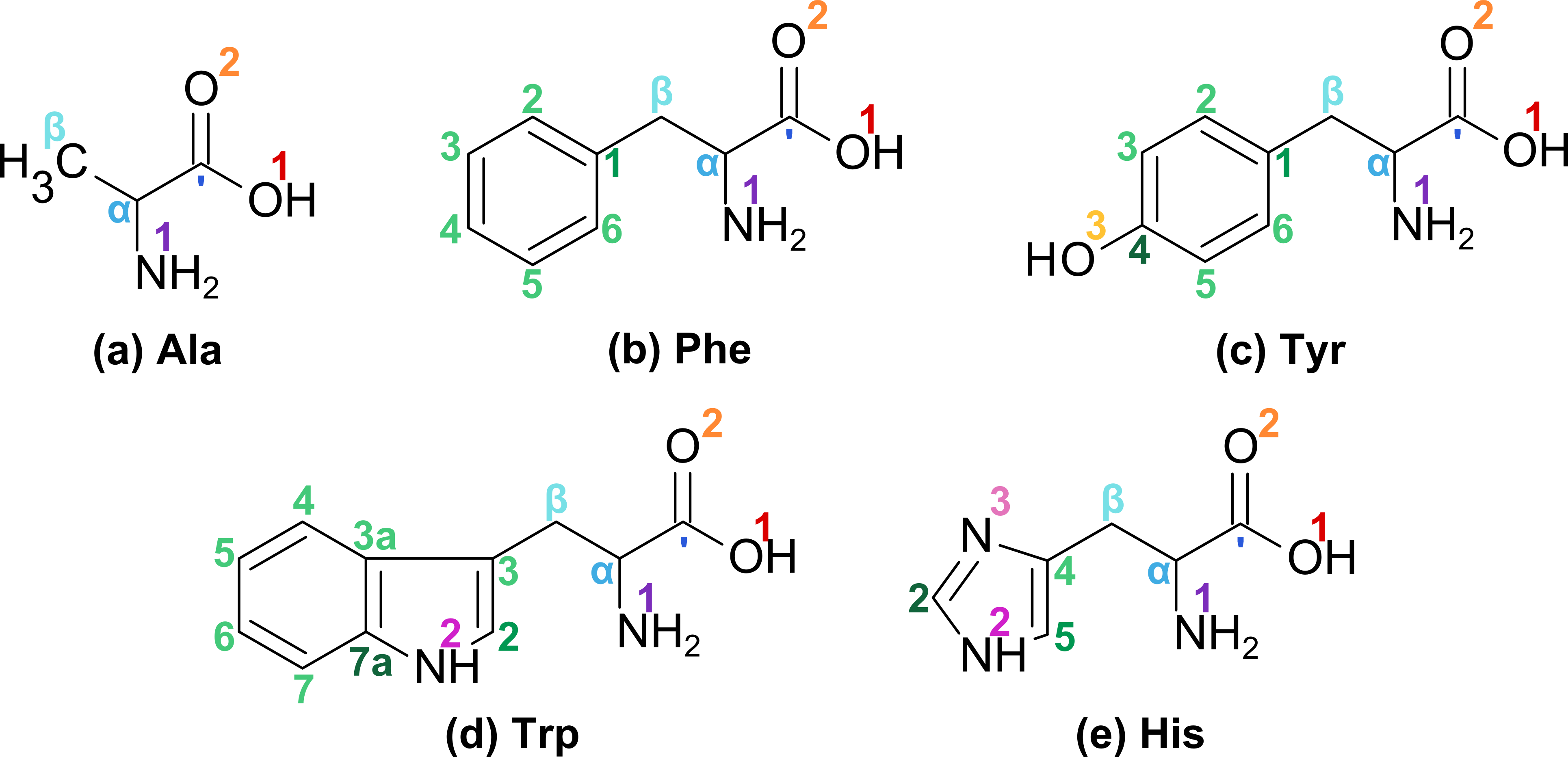}
\caption{Schematic of (a) Ala, (b) Phe, (c) Tyr, (d) Trp and (e) His showing the atomic structures and atom labels, which will be used in the following.   \label{fig:aa_schem}}
\end{figure}

Amino acids form the basis of peptides and proteins, which are fundamental building blocks of life, and they are of great scientific interest for a multitude of reasons, first and foremost due to their role in biology and related use in pharmacology and medicine. Their systematic nature also makes them perfect test systems to understand important aspects of the behaviour of molecular systems, including local and long-range structure and interactions, polymorphism, the three dimensional arrangement of proteins, and ionic behaviour and its tunability by the environment. Whilst the motivation to study amino acids is clear, experimental strategies are generally limited to structural techniques such as X-ray diffraction (XRD). A complementary technique, which can provide an additional level of information on chemical states and electronic structure not accessible to XRD, is X-ray photoelectron spectroscopy (XPS). Recently, we have started to explore the application of XPS to amino acids in their crystalline, powder form in combination with theoretical calculations based on density functional theory (DFT)~\cite{Hohenberg1964,Kohn1965}. Our first study established a combined experiment-theory approach to predict and interpret primarily the C~1\textit{s} core level spectra of the simple amino acids glycine (Gly), alanine (Ala), and serine (Ser)~\cite{Pi2020}. Here, we expand and improve our previous approach to amino acids with aromatic side chains, including phenylalanine (Phe), tyrosine (Tyr), tryptophan (Trp), and histidine (His). Fig.~\ref{fig:aa_schem} shows a schematic of their atomic structures, including Alanine (Ala) which is used as a reference throughout and which we have reported previously\cite{Pi2020}.\\

As for X-ray photoelectron spectroscopy studies on amino acids in general, very few studies exist on the aromatic subgroup. A small number of experiments have been performed on Phe, Tyr and His adsorbed on single crystal substrates, including Au, Ag, Cu and TiO\textsubscript{2}\cite{Thomas2007AdsorptionSurface,Feyer2008,Feyer2010AdsorptionAu111,Reichert2010L-tyrosineScheme}. Whilst gas phase experiments are often used to study amino acids, this is difficult to achieve for the aromatic subgroup as they generally have high melting points (and consequently low vapour pressures) as well as low thermal stability\cite{Zhang2009ElectronicSpectroscopy}.
A very limited number of studies on solid powders has been reported, which often suffer from low experimental resolution complicating peak assignments\cite{Zubavichus2004SoftStudy,Cardenas2006TheSpectroscopy}. The 2013 work by Stevens \textit{et al.}\ provides the most systematic and detailed study of solid phase amino acids to date, in which only His of the aromatic subgroup is included\cite{Stevens2013}. Beyond XPS, X-ray absorption spectroscopy and electron energy loss spectroscopy have been employed to understand the chemistry and structure of the aromatic amino acids\cite{Boese1997CarbonPeptides,Cooper2004InnerPhenylalanine}. Due to the complexity of aromatic amino acids, the use of theory to guide the interpretation of spectra is essential. This is particularly true in the solid state, where intermolecular interactions can have an important effect, posing a further challenge to peak assignment. 
Nonetheless, from a theoretical point of view, only a handful of examples of core binding energy (BE) calculations of aromatic amino acids exist, which are all limited to the gas phase~\cite{Wang2008,Ganesan2009,Zhang2009ElectronicSpectroscopy,Ganesan2014,Wang2014}. Furthermore, to the best of our knowledge the core state BEs of His have not previously been calculated.\\

In this work, the aromatic amino acids phenylalanine (Phe), tyrosine (Tyr), tryptophan (Trp) and histidine (His) are explored using both experiment and theory. The subgroup classification of amino acids usually includes Phe, Tyr and Trp in the aromatic group with Tyr also sometimes grouped with the polar amino acids. Due to the basic properties of His it is often classified as a polar amino acid. For completeness, we include all amino acids containing aromatic side chains here, independent of their polar nature. XPS experiments in the solid phase are compared to theoretical calculations based on DFT using the $\Delta$SCF (self-consistent field) approach as implemented in systematic basis sets. Due to the complexity of the observed core level spectra and the apparent strong influence of not only nearest, but also next-nearest and even further removed neighbouring atoms, a molecular subspecies approach was followed to aid in the rationalisation and explanation of observed BE shifts, particularly for the C and N~1\textit{s} core states. This is shown to be an extremely useful approach to gain a full and detailed understanding of the chemical and electronic structure of these important biological building blocks.\\

\section{Methods}

\subsection{Theoretical Approach}

Density functional theory was used to calculate the core state BEs of Ala, Phe, Tyr, Trp, and His.  The primary motivation is the calculation of solid state BEs to aid the interpretation of experimental spectra.  However, the solid state BEs are influenced by a combination of factors including the presence of different functional groups, the molecular and crystal structure, and intermolecular interactions.  Theory is essential to disentangle these competing effects. Initial crystal structures for Ala, Phe, Tyr, Trp and His were obtained from Refs.~\cite{Destro1988,Ihlefeldt2014,Mostad1972,Gorbitz2015,Fronczek2016}, respectively. In order to assess the influence of the molecular structure, different gas phase conformers were tested for each molecule and selected as follows. Four low energy conformers for Phe, Tyr and Trp, and three low energy conformers for His were taken from the literature~\cite{Purushotham2012,Ropo2016}. Following geometry optimisation and BE calculations, the two conformers of each amino acid with the most distinct BEs were retained for further investigation.  For Ala, only the lowest energy conformer used in Ref.~\cite{Pi2020} was considered. The  structures of each conformer are presented in the Supplementary Information. The gas phase conformers were compared to the molecule extracted directly from the bulk, which was allowed to relax away from the zwitterionic state into its neutral form. BEs were calculated using the $\Delta$SCF approach, however, in order to distinguish between initial and final state effects, gas phase BEs were additionally calculated at the level of Koopmans'. To determine the contribution from intermolecular interactions, the BEs of both the gas and solid phase are compared. Finally, in order to assess the impact of different functional groups, a systematic series of subspecies molecules was investigated, as depicted in Fig.~\ref{fig:all_molecules}, which are derived from the aromatic amino acids studied here.\\

In order to aid interpretation of experimental spectra, the relative BE positions of contributing chemical environments are needed. Absolute BEs are not necessary for this approach, and DFT is more reliable for relative than absolute BEs. However, some recent work exists where DFT has been shown to accurately reproduce absolute BEs~\cite{Kahk2019,Ozaki2017}. When comparing calculations across the molecules, it is important to note that while BEs calculated for molecules in the gas phase can be directly compared between molecules, this is not the case for solid state calculations, since the core hole calculations are performed in charged supercells. Although schemes exist to account for the use of periodic boundary conditions (e.g.\ Ref.~\cite{Ozaki2017}), this can introduce an additional source of uncertainty. Therefore, since it is not essential for the current work, BEs of the amino acids in the solid state are not directly compared.\\

\begin{landscape}
\vspace*{\fill}
\begin{figure*}
\centering
\includegraphics[width=\paperwidth]{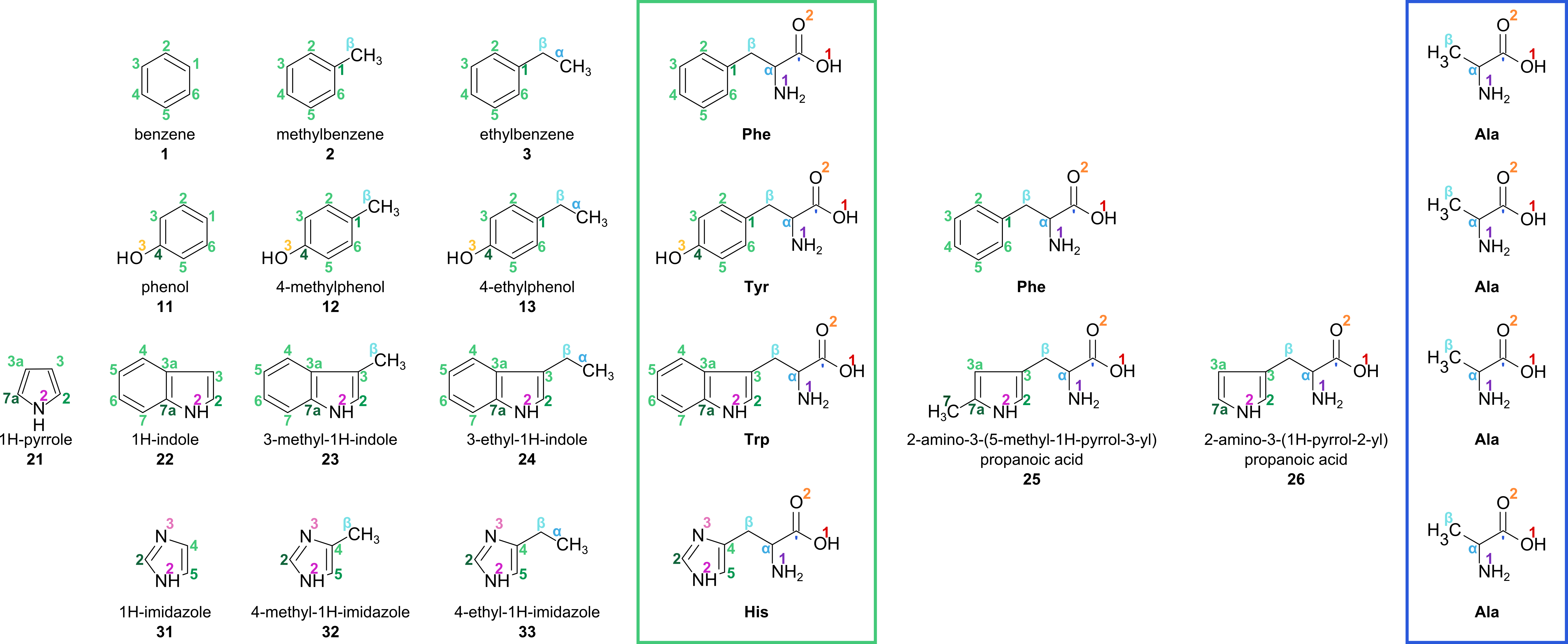}
\caption{Molecular structure of the amino acids Ala, Phe, Tyr, Trp and His, as well as the molecular subspecies calculated, including atom labels used throughout this work. The molecules included are (\textbf{1}) benzene, (\textbf{2}) methylbenzene, (\textbf{3}) ethylbenzene, (\textbf{11}) phenol, (\textbf{12}) 4-methylphenol, (\textbf{13}) 4-ethylphenol, (\textbf{21}) 1H-pyrrole, (\textbf{22}) 1H-indole, (\textbf{23}) 3-methyl-1H-indole, (\textbf{24}) 3-ethyl-1H-indole, (\textbf{25}) 2-amino-3-(5-methyl-1H-pyrrol-3-yl)propanoic acid, (\textbf{26}) 2-amino-3-(1H-pyrrol-2-yl)propanoic acid, (\textbf{31}) 1H-imidazole, (\textbf{32}) 4-methyl-1H-imidazole, and (\textbf{33}) 4-ethyl-1H-imidazole.\label{fig:all_molecules}}
\end{figure*}
\vfill
\end{landscape}

\subsubsection{Computational Details}

Gas phase geometry optimisations were performed using BigDFT~\cite{Genovese2008,Ratcliff2020}, in open boundary conditions, with a wavelet grid spacing of 0.185~\AA, coarse and fine radius multipliers of 5 and 8, respectively, and HGH-GTH pseudopotentials (PSPs)~\cite{Goedecker1996,Hartwigsen1998}.
Gas phase BE calculations were performed at the level of both Koopmans' and $\Delta$SCF using the MADNESS molecular DFT code~\cite{Harrison2016} with open boundary conditions. A mixed all-electron (AE)/PSP approach was used~\cite{Ratcliff2019}, wherein the atom of interest was treated at the AE level, with remaining atoms treated at the PSP level, as described in Ref.~\cite{Pi2020}. 
Ground state calculations used a wavelet threshold of $10^{-4}$ followed by $10^{-6}$ (wavelet order $k=6$ and $k=8$), while core hole
calculations directly used a wavelet threshold of $10^{-6}$ ($k=8$). A convergence criterion of $10^{-3}$ was
used for both the density and Kohn-Sham wavefunction residuals. 
Following Ref.~\cite{Pi2020} the ground state wavefunctions were used as an input guess for the core hole calculations, localisation was imposed on the 
wavefunctions for the ground state while core hole calculations used canonical orbitals, and the B-spline projection based derivative operator was used (except for the
calculation of the kinetic energy operator)~\cite{Anderson2019}. Calculations employed the same PSPs as BigDFT.\\

Solid state geometry optimisations and BE calculations using the $\Delta$SCF approach were performed with the CASTEP plane-wave DFT code~\cite{Clark2005}. Core hole PSPs were used to represent the core-excited atom, following the same procedure and with the same norm-conserving on-the-fly generated PSPs as Ref.~\cite{Pi2020}. 
Calculations were performed with a cut-off energy of 900~eV, and Monkhorst-Pack~\cite{Monkhorst1976} $k$-point grids of $2 \times 1 \times 2$, $2 \times 2 \times 1$, $2 \times 1 \times 2$, and $2 \times 2 \times 2$ for Ala, Phe, Tyr and His, respectively, with Trp calculations performed at the $\Gamma$-point only.
Geometry optimisations used the semi-empirical dispersion correction scheme of Grimme~\cite{Grimme2006}. \\

Gas phase BE calculations were performed using PBE only~\cite{Perdew1996}, while solid state BE calculations were performed using both PBE and PBE0~\cite{Adamo1999}, except for Phe and Trp where PBE0 calculations were prohibitively expensive due to their large unit cells containing 184 and 432 atoms, respectively.  All BEs were calculated in the vertical approximation.\\ 

All geometry optimisations used the PBE functional and a force tolerance of 0.02~eV/\AA.  For solid state geometry optimisations the cell was also allowed to relax.
For molecules extracted from the optimised crystals, only the H atoms were relaxed, with all other atoms frozen. In order to prevent collapse back to the zwitterionic state, an initial perturbation was applied to one of the H atoms.  For all other gas phase calculations, all atoms were allowed to relax. All calculations were spin restricted and relativistic effects were neglected, since although these can have a significant effect when calculating absolute BEs (see e.g.\ Ref.~\cite{Besley2009}), they are less significant when considering relative BEs, as in this work.
The same computational parameters were used for both gas phase amino acids and molecular subspecies calculations.
Molecule and crystal structures were visualized using VESTA~\cite{Momma2008}.

\subsection{Experimental Approach}

Powders of the L-stereoisomers of all investigated amino acids were purchased from Sigma-Aldrich (Ala~$\geqslant$99\%, Phe~$\geqslant$98\%, Tyr~$\geqslant$98\%, Trp~$\geqslant$98\%, His~$\geqslant$99\%). Core level spectra were recorded on a Thermo Scientific K-Alpha+ XPS system with a monochromated, microfocused Al K$\alpha$ X-ray source (h$\nu$ = 1486.7~eV), which was operated a 6~mA emission current and 12~kV anode bias. The base pressure was 2$\times10^{-9}$ mbar. All core level spectra were collected at a pass energy of 20~eV using an X-ray spot size of 400~$\mu$m. Samples were mounted on conducting carbon tape and a flood gun was employed to prevent sample charging. As amino acids are prone to suffer from radiation damage, samples were rastered and data collected at four points across the samples, which were then averaged to achieve the necessary signal statistics for peak fitting. All data were analysed using the Avantage software package. Differences in peak positions across the different measurement points were less than 50~meV for all core levels. For peak fit analysis, Shirley-type backgrounds and Voigt functions were used with both the full width at half maximum (FWHM) and Lorentzian/Gaussian (L/G) ratios refined.\\

\section{Results}

\subsection{Calculated Solid State Binding Energies}

In line with our previous work~\cite{Pi2020}, the use of PBE with semi-empirical dispersion corrections for the solid state geometry optimisations resulted in a good description of the crystal structure.   Relaxed lattice parameters and angles, which are reported in the Supplementary Information alongside the relaxed crystal structures, are in good agreement with the experimental values, with maximum discrepancies of 3.0~\% and 1.2~\%, respectively.\\

Calculated BEs for the solid state amino acids are presented in Tab.~\ref{tab:bes}.
Due to the relatively large unit cell sizes of the aromatic amino acids, it is highly desirable to perform calculations using semi-local functionals such as PBE, rather than hybrid functionals such as PBE0. Indeed, for Phe and Trp PBE0 BE calculations were prohibitively expensive. For the amino acids where PBE0 calculations were possible (Ala, Tyr, and His), significant quantitative differences can be seen between the two functionals for C~1\textit{s}, up to 0.7~eV in the most severe cases. However, qualitatively the differences are less significant, and it is primarily the BE of C$'$ relative to the other states which is most strongly affected. Importantly, the order of BEs remains constant to within 0.1~eV, so that for the purposes of aiding in peak assignment in experimental spectra it is not necessary to go beyond PBE. Furthermore, the differences for O and N~1\textit{s} core states are negligible.  Therefore, the calculations presented in the following sections were performed using PBE only.

\begin{table*}[h]
\centering
\begin{threeparttable}
\caption{\label{tab:bes} Experimental (`Exp') and PBE/PBE0 calculated relative C, O and N 1\textit{s} core state BEs for solid state amino acids.}
\begin{tabular*} {1.0\textwidth}{c @{\extracolsep{\fill}} rrr rr rrr rr rrr}

\hline\hline

 & \multicolumn{3}{c}{Ala} & \multicolumn{2}{c}{Phe} & \multicolumn{3}{c}{Tyr} & \multicolumn{2}{c}{Trp} & \multicolumn{3}{c}{His} \\
 

 

& PBE & PBE0  & Exp & PBE &  Exp & PBE & PBE0  & Exp & PBE   & Exp & PBE & PBE0  & Exp \\
 

\cline{1-1} \cline{2-4} \cline{5-6} \cline{7-9} \cline{10-11} \cline{12-14}  \\[-2.5ex]
 \\[-2.5ex]

C$'$ & 0.0 & 0.0 & 0.0 &
0.0 & 0.0 &
0.0 & 0.0 & 0.0 &
0.0 & 0.0 &
0.0 & 0.0 & 0.0 \\

C$^\alpha$ & -1.7 & -2.1 & -2.0 &
-1.4 & -1.9 & 
-1.5 & -2.1 & -2.5 & 
-1.3 & -2.1 & 
-1.8 & -2.1 & -1.9\\

C$^\beta$ & -3.1 & -3.5 & -3.3 &
-2.5 & -2.6 & 
-2.7 & -3.4 & -3.6 & 
-2.3 & -3.1 & 
-2.7 & -3.2 & -3.2 \\

C$_1$ & - & - & - &
-2.9 & -3.1 & 
-3.6 & -4.3 & -4.3 & 
- & - &
- & - & - \\

C$_2$ & - & - & - &
-3.2 & -3.7 & 
-3.7 & -4.4 & -4.3 &
-2.7 & -3.7 & 
-2.1 & -2.5 & -2.4\\

C$_3$ & - & - & - &
-3.1 & -3.7 & 
-3.8 & -4.5 & -4.3 & 
-3.3 & -4.3 & 
- & - & - \\

C$_{3\mathrm{a}}$ & - & - & - &
- & - &
- & - & - &
-3.2 & -4.3 & 
- & - & - \\

C$_4$ & - &	- & - &
-3.2 & -3.7 & 
-2.1 & -2.8 & -3.1 &
-3.4 & -4.3 &
-2.8 & -3.2 &  -3.2 \\

C$_5$ & - & - & - &
-3.2 & -3.7 & 
-3.8 & -4.5 & -4.3 & 
-3.4 & -4.3 & 
-2.8 & -3.3 & -3.2\\  

C$_6$ & - & - & - &
-3.1 & -3.7 & 
-3.6 & -4.3 & -4.3 & 
-3.3 & -4.3 & 
- & - & - \\

C$_7$ & - & - & - &
- & - & 
- & - & - &
-3.2 & -4.3 & 
- & - & - \\

C$_{7\mathrm{a}}$ & - & - & - &
- & - & 
- & - & - &
-2.3 & -3.1 & 
- & - & - \\

\cline{1-1} \cline{2-4} \cline{5-6} \cline{7-9} \cline{10-11} \cline{12-14}  \\[-2.5ex]

O$^1$ & 0.0 & 0.0 & 0.0 &
0.0 & 0.0 & 
0.0 & 0.0 & 0.0 &
0.0 & 0.0 & 
0.0 & 0.0 & 0.0 \\

O$^2$ & 0.2 & 0.2 & 0.0 & 
0.0	& 0.0 & 
0.1 & 0.1 & 0.0	& 
-0.1 & 0.0	& 
0.0 & 0.0 & 0.0 \\

O$^3$ & - & - & - &
- & - & 
0.8 & 0.8 & 1.2 &
- & - & 
- & - & \\

\cline{1-1} \cline{2-4} \cline{5-6} \cline{7-9} \cline{10-11} \cline{12-14}  \\[-2.5ex]

N$^1$ & 0.0 & 0.0 & 0.0 &
0.0 & 0.0 &
0.0 & 0.0 & 0.0 &
0.0 & 0.0 &
0.0 & 0.0 & 0.0 \\

N$^2$ & - & - & - &
- & - &
- & - & - &
-1.2 & -1.5 & 
-0.6 & -0.7 & -1.0\\

N$^3$ & - & - & - &
- & - &
- & - & - &
- & - &
-2.2 & -2.3 & -2.7  \\

\hline\hline
\end{tabular*}

\end{threeparttable}
\end{table*}

Whilst the calculated BE positions describe the experimental core level spectra very well, which will be discussed in detail in Section~\ref{sec:spectra}, it is not easy to intuitively rationalise the order and relative positions of the different constituents, in particular for the case of C~1\textit{s} with its many chemical states. Therefore, a molecular subspecies approach was chosen to systematically explore core level energy changes with the removal or introduction of part of the amino acids and their functional groups.\\

\subsection{Molecular Subspecies Series}

\begin{figure*}[p]
\centering
\includegraphics[scale=0.25]{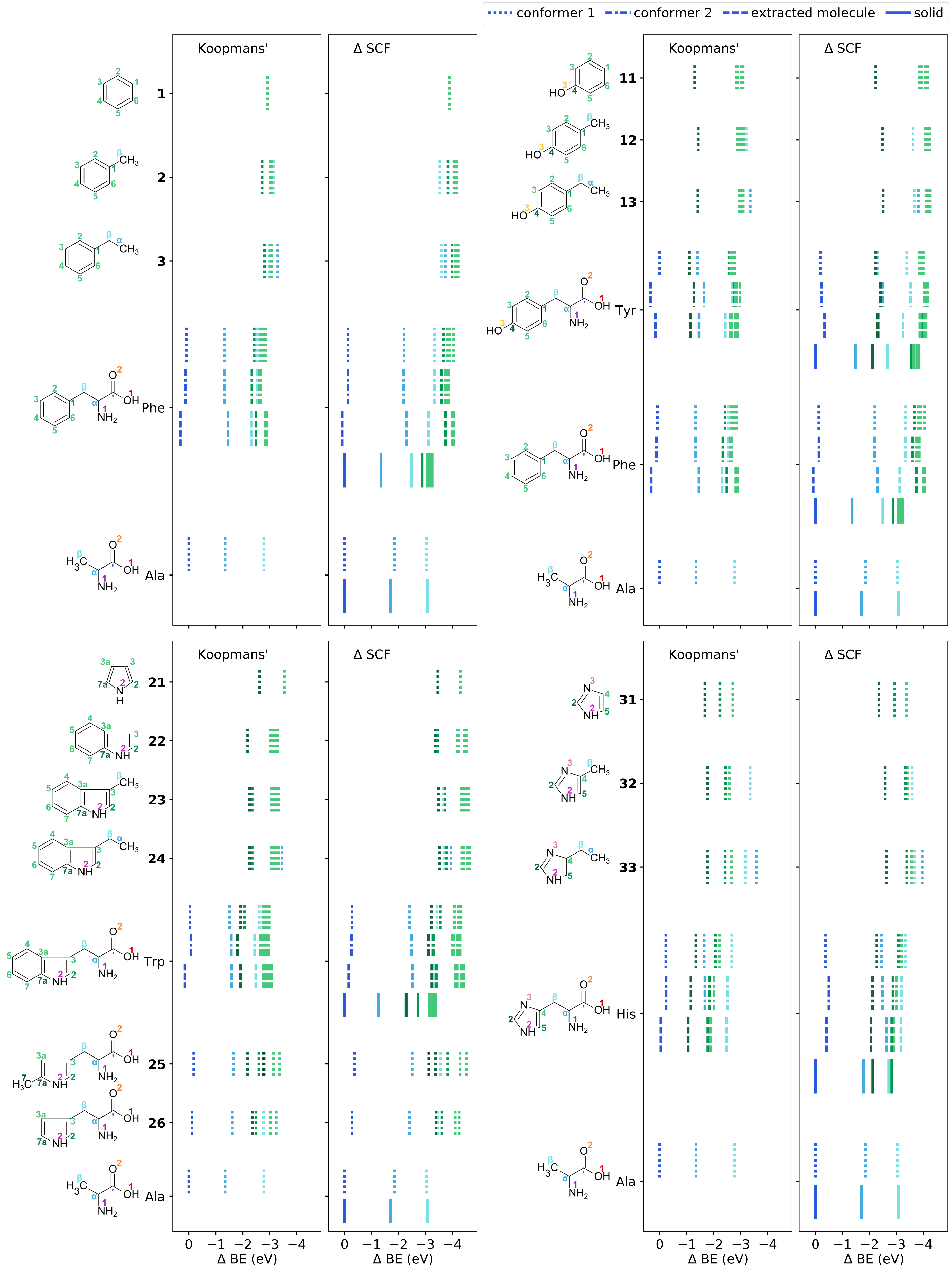}
\caption{PBE-calculated C~1\textit{s} BEs for the amino acids and the series of subspecies molecules. Gas phase BEs are relative to Ala C$'$, while solid state BEs are relative to C$'$ of that amino acid.  \label{fig:mol_series}}
\end{figure*}

Twenty-two additional small molecular systems were explored theoretically to aid our understanding of the core level spectra observed for the aromatic amino acids. Fig.~\ref{fig:all_molecules} gives an overview of the main set of molecular subspecies calculated and their relationship to the aromatic amino acids and Ala. Fig.~\ref{fig:mol_series} provides an overview of the C~1\textit{s} BEs of the molecular series, while the tables of the corresponding BEs are also given in the Supplementary Information. A set of additional subspecies was explored to understand specific questions arising around nitrogen groups and aromatic systems, which is shown in the Supplementary Information. In the following subsections the results and main conclusions for each of the aromatic amino acids are discussed.\\

\subsubsection{Phe}

In parallel to Phe being the simplest of the amino acids explored here, it also reduces to the simplest submolecule, benzene (\textbf{1}). As expected, all C atoms for benzene have the same BEs as each other in both the Koopmans' and the $\Delta$SCF approaches. Moving to methylbenzene (\textbf{2}) a clear difference between C$_1$ and the remaining C atoms of the aromatic ring is noticeable, in line with previous calculations~\cite{Myrseth2007}.
This is clearly illustrated by the differences between the ground state electronic densities of (\textbf{2}) and  (\textbf{1}), which are depicted in the Supplementary Information, where the addition of the CH\textsubscript{3} group changes the density around C$_1$.  There are also non-negligible changes in the density around all other aromatic C atoms C$_{\mathrm{arom}}$. 
Combined with changes in the atomic structure between (\textbf{2}) and  (\textbf{1}), and which are not accounted for in the visualisation of the densities, this explains why the Koopmans' BEs of all C atoms change between the two molecules.
Comparing to experimental gas phase measurements by Ohta \textit{et al.}~\cite{Ohta1975Core-electronSpectroscopy}, we observe that while the relative BEs agree reasonably well with experiment, their peak assignments are more in line with the Koopmans' values. In particular, C$_1$ has the highest BE, while C$^\beta$ is at the lowest BE.\\

Whilst C$^\beta$ in (\textbf{2}) and ethylbenzene (\textbf{3}) and  C$^\alpha$ in (\textbf{3}) occur at the lowest BEs, this changes completely using the $\Delta$SCF approach, where C$^\alpha$ and C$^\beta$ move to the higher BE side of all other C atoms.
In addition, a clear chemical shift between the CH\textsubscript{2} and CH\textsubscript{3} groups of the side chain for (\textbf{3}) is also apparent.
In order to understand to what extent the shift in C$^\beta$ for $\Delta$SCF is affected by the aromaticity of (\textbf{2}), we also compare with methylcyclohexane (\textbf{44}), for which results are given in the Supplementary Information. In particular, the $\Delta$SCF results for (\textbf{44}) only show a small spread, but otherwise both C$^\beta$ and C$_1$ are at very similar energies to the remaining C atoms, in contrast to (\textbf{2}).  In other words, the conjugated system is much more sensitive to the addition of the CH\textsubscript{3} group when final state effects are taken into account.
When examining the density difference between (\textbf{3}) and  (\textbf{2}), there is a small change in the density around C$_1$, which in turn gives rise to a small change in the Koopmans' BEs.  All other C atoms in the ring, however, remain unaffected by the addition of the CH\textsubscript{3} group, so that the corresponding BEs of the C$_{\mathrm{arom}}$ atoms do not change between (\textbf{2}) and (\textbf{3}).\\

For Phe itself the C$_{\mathrm{arom}}$ atoms including C$_1$ behave similarly to (\textbf{1})-(\textbf{3}), with a clear spreading in BE of C$_1$-C$_6$. With the addition of the carboxylic COO\textsuperscript{-} group the separation between C$_{\mathrm{arom}}$ and C$^\alpha$ and C$^\beta$ increases significantly and C$^\alpha$ and C$^\beta$ switch places. 
The $\Delta$SCF results for the gas phase molecules show similar variations between conformers and are in agreement with previous calculations from Zhang \textit{et al.}~\cite{Zhang2009ElectronicSpectroscopy}\ who included four different conformers. When comparing the $\Delta$SCF gas phase conformers with the solid Phe a clear bunching up of BEs is observed, whilst the relative BE order of the different C environments remains the same. The significant change in BE of C$^\alpha$ and C$'$ can be explained by the change from COOH/NH\textsubscript{2} to the zwitterionic COO\textsuperscript{-}/NH\textsubscript{3}\textsuperscript{+} environments and the resulting intermolecular interactions. The main observable difference between the Koopmans' and $\Delta$SCF results for Phe lies in the differentiation of C$^\beta$ from C$_{\mathrm{arom}}$.  Whilst they are very close in energy or even overlap for some conformers, C$^\beta$ moves to significantly higher BEs in $\Delta$SCF due to final state effects, which is consistent with the behaviour of C$^\beta$ in (\textbf{2}) and (\textbf{3}).\\

\subsubsection{Tyr}

Across the series from phenol (\textbf{11}) to 4-methylphenol (\textbf{12}) and 4-ethylphenol (\textbf{13}) a common feature is the spreading out of C$_{\mathrm{arom}}$ BEs due to the presence of the hydroxyl group.
Similarly to the equivalent series for Phe, there is a significant change in the electronic density (shown in the Supplementary Information) on all C$_{\mathrm{arom}}$ going from (\textbf{11}) to (\textbf{12}), with corresponding changes in the BEs. However, the changes in density between (\textbf{12}) and (\textbf{13}) are again primarily localized on C$_1$, with the remaining C$_{\mathrm{arom}}$  unaffected by the addition of the CH\textsubscript{3} group.
In parallel to the spreading out of the C$_{\mathrm{arom}}$ BEs, a large gap also opens up between C$_4$ and the remaining C atoms. Comparing (\textbf{11}) with cyclohexanol (\textbf{47}), for which results are presented in the Supporting Information, this gap is much larger in (\textbf{11}) than in (\textbf{47}) for both Koopmans' and $\Delta$SCF, demonstrating the strong influence of the aromaticity and the importance of final state effects. Both Koopmans' and $\Delta$SCF results for (\textbf{11}) agree well with experimental gas phase results from Ohta \textit{et al.}~\cite{Ohta1975Core-electronSpectroscopy}. One interesting point to note about (\textbf{11}) is the difference in BEs between C$_3$ and C$_5$, and  C$_2$ and C$_6$, which in contrast have the same BEs in both (\textbf{2}) and  aniline (\textbf{46}) and the same is true for the equivalent non-conjugated molecules. However, due to the presence of the hydroxyl group, neither (\textbf{11}) nor (\textbf{47}) have symmetric structures, and the small asymmetry of the BEs can be attributed to this asymmetry of the atomic structures. \\

As was the case for the subspecies molecules for Phe, C$^\alpha$ and C$^\beta$ in (\textbf{12}) and (\textbf{13}) occur at the lowest BEs in the Koopmans' approach, but swap when $\Delta$SCF is used. Compared to Phe, C$^\beta$ in Tyr shifts to even higher BE relative to C$_{\mathrm{arom}}$ in the $\Delta$SCF approach. This is a direct result of the addition of the hydroxyl group onto the aromatic ring and showcases the strong long-range intramolecular interactions taking place. Of course C$_4$ is now also clearly separated from the rest of the aromatic ring and located at a BE intermediate between C$^\alpha$ and C$^\beta$.\\

As with Phe, there is also variation between Tyr conformers, where the results are again in line with calculations from Zhang \textit{et al.}~\cite{Zhang2009ElectronicSpectroscopy}. A significant change in the BE separation of C$^\alpha$ and C$_4$ occurs when moving from the gas phase calculations to the solid state case. Whilst in the gas phase their binding energies are almost identical across all Tyr molecules considered, they separate significantly in the solid. This is due to the hydroxyl group taking part in intermolecular hydrogen bonding as can be clearly seen from the crystal structures shown in the Supplementary Information.\\

\subsubsection{Trp}

1H-pyrrole (\textbf{21}) nicely exemplifies the symmetric nature of the ring with C$_2$/C$_{7a}$ and C$_3$/C$_{3a}$ grouping together for both Koopmans' and $\Delta$SCF. In 1H-indole (\textbf{22}) C$_2$ and C$_{7a}$ remain at significantly higher BEs than all other C atoms. Comparing the Koopmans' and $\Delta$SCF results for 3-methyl-1H-indole (\textbf{23}) and 3-ethyl-1H-indole (\textbf{24}) a considerable change in BE for C$^\alpha$ and C$^\beta$ is observed as in the previous cases discussed. A systematic difference in the $\Delta$SCF BEs of C$_2$ and C$_{7a}$ is noted across all molecules in the series except (\textbf{21}), even if the six-membered ring is removed as is the case in 2-amino-3-(5-methyl-1H-pyrrol-3-yl)propanoic acid (\textbf{25}) and 2-amino-3-(1H-pyrrol-2-yl)propanoic acid (\textbf{26}).\\

As with Phe and Tyr, and again in agreement with Zhang \textit{et al.}~\cite{Zhang2009ElectronicSpectroscopy}, there is noticeable variation between the Trp conformers.  While the Koopmans' BEs for Trp are in line with chemical intuition, the $\Delta$SCF BEs are harder to explain. In particular, contrary to the expectation that aromatic and aliphatic C atoms should have similar BEs, C$^\beta$ is noticeably higher in BE than the C$_{\mathrm{arom}}$ which do not neighbour a N atom.  Indeed, the BE of C$^\beta$ is especially sensitive to final state effects, as evidenced by the difference between Koopmans' and $\Delta$SCF values.
This is also the case for both Phe and Tyr, and by comparing (\textbf{2}) and (\textbf{44}) was attributed to the conjugated nature of the ring. Similarly, the density comparisons discussed in relation to Phe and Tyr demonstrated that the functionalisation of an aromatic ring can impact on the density and thus the BEs of \textit{all} atoms in the ring, not just the nearest neighbour.  This explains for example why it is not just the BE of C$^\alpha$ which is affected by the addition of the amino group when going from (\textbf{24}) to Trp.\\

Furthermore, in Trp the BE of C$^\beta$ is surprisingly close to that of both C$_2$ and C$_{7\mathrm{a}}$ in the gas phase and the same as C$_{7\mathrm{a}}$ in the solid state, which cannot be explained by arguments based purely on electronegativity. On the contrary, since they each neighbour a N atom, one would expect the BE of C$^\alpha$ to be close to that of C$_2$ and C$_{7\mathrm{a}}$, which is not the case in either Trp, (\textbf{25}), or (\textbf{26}). In addition to next-nearest neighbour effects, this can also be explained by the protonation state of the N atoms. In order to provide further insights on the influence of different protonation states of N on C~1\textit{s} BEs, we also considered an additional set of subspecies molecules containing nitrogen, for which results are given in the Supplementary Information. Taking for example the series of ethylamine (\textbf{41}) to diethylamine (\textbf{42}) to triethylamine (\textbf{43}), one can see a clear trend in the $\Delta$SCF BEs, where the higher the protonation state of the N atom, the higher the BE of C$^\alpha$. This trend is in agreement with C$^\alpha$ having a higher BE than C$_2$ and C$_{7\mathrm{a}}$. Finally, we note that the BEs of C$^\beta$ in the alkylamine series are also affected by the change in N protonation state, providing further support for the importance of next-nearest neighbour effects, although the magnitude of variations is much smaller than for C$^\alpha$.\\

To further test the influence of aromaticity, the BEs for (\textbf{46}) and cyclohexanamine (\textbf{45}) were calculated, for which results are given in the Supplementary Information.  Both Koopmans' and $\Delta$SCF results for (\textbf{46}) are in good agreement with experimental gas phase results from Ohta \textit{et al.}~\cite{Ohta1975Core-electronSpectroscopy}.  Consistent with (\textbf{11}) and (\textbf{47}), a larger gap between C$_1$ and the remaining C atoms is observed for (\textbf{46}) than (\textbf{45}), while there is also a larger spread of the C atoms in the ring in  (\textbf{46}) compared to (\textbf{45}). A clear overall trend is observed upon the addition of a functional group to a ring, whether conjugated and non-conjugated, where an increasing split between the C atom the group binds to and the remaining C atoms is observed in line with the increasing electronegativity in going from C to N to O in the functional groups CH\textsubscript{3}, NH\textsubscript{2}, and OH. Comparing the effect on the conjugated versus non-conjugated rings, this difference is always bigger for the conjugated ring.\\

\subsubsection{His}

In 1H-imidazole (\textbf{31}) the three C atoms all have considerably different BEs, including a clear distinction in C BE depending on the protonation of the neighbouring N atom in line with the previous observations for molecules (\textbf{41})-(\textbf{43}). The addition of the methyl and ethyl side chains in 4-methyl-1H-imidazole (\textbf{32}) and 4-ethyl-1H-imidazole (\textbf{33}), respectively, reduces the difference in BE between C$_4$ and C$_5$. Going from Koopmans' to $\Delta$SCF a significant change in the BEs of C$^\beta$ and C$^\alpha$ relative to the three C atoms in the aromatic ring, C$_2$, C$_4$, and C$_5$, is observed. The relative differences between C$_2$, C$_4$, and C$_5$ remain very similar between the two approaches.\\  

Across all $\Delta$SCF gas phase calculations of His, C$_4$, C$_5$ and C$^\beta$ are very close in BE. This is comparable to the observations made for C$_2$, C$_{7\mathrm{a}}$ and C$^\beta$ in Trp. Another similarity between Trp and His is that C$^\alpha$ is the most sensitive to changes in conformer and gas/solid phases, and its BE changes significantly between calculations. In the solid phase C$^\alpha$ is even higher in BE than C$_2$, which is not the case in either the Koopmans' or $\Delta$SCF gas phase calculations, and this is most certainly not immediately intuitive. However, based on the results presented so far, this is a consequence of a complex interplay between the protonation of the N atoms, the influence of the aromatic ring, and the intermolecular interactions of both the NH\textsubscript{3}\textsuperscript{+} and NH groups in His. As will be discussed in more detail in the following section, previous experimental work by Stevens \textit{et al.}\ assigned the chemical states present closely to the results we find for the Koopmans' approach \cite{Stevens2013}.\\ 

To summarise the observations made to this point, the molecular subspecies approach is invaluable to rationalise and discuss the complex relative BE changes observed in the amino acids. A fascinating, if somewhat subjective, result from the combination of experiment and theory and the exploration of the molecular subspecies is that chemical intuition and the experience of a spectroscopist usually reflects the results given by Koopmans' theorem. The additional rearrangement of BE positions observed in $\Delta$SCF is often surprising, resulting in our hypothesis that human brains are not best placed to compute final state effects ad hoc without the aid of DFT.\\

\subsection{Core Level Spectra of the Amino Acids}\label{sec:spectra}

\begin{figure*}[ht]
\centering
\includegraphics[width=1.0\textwidth]{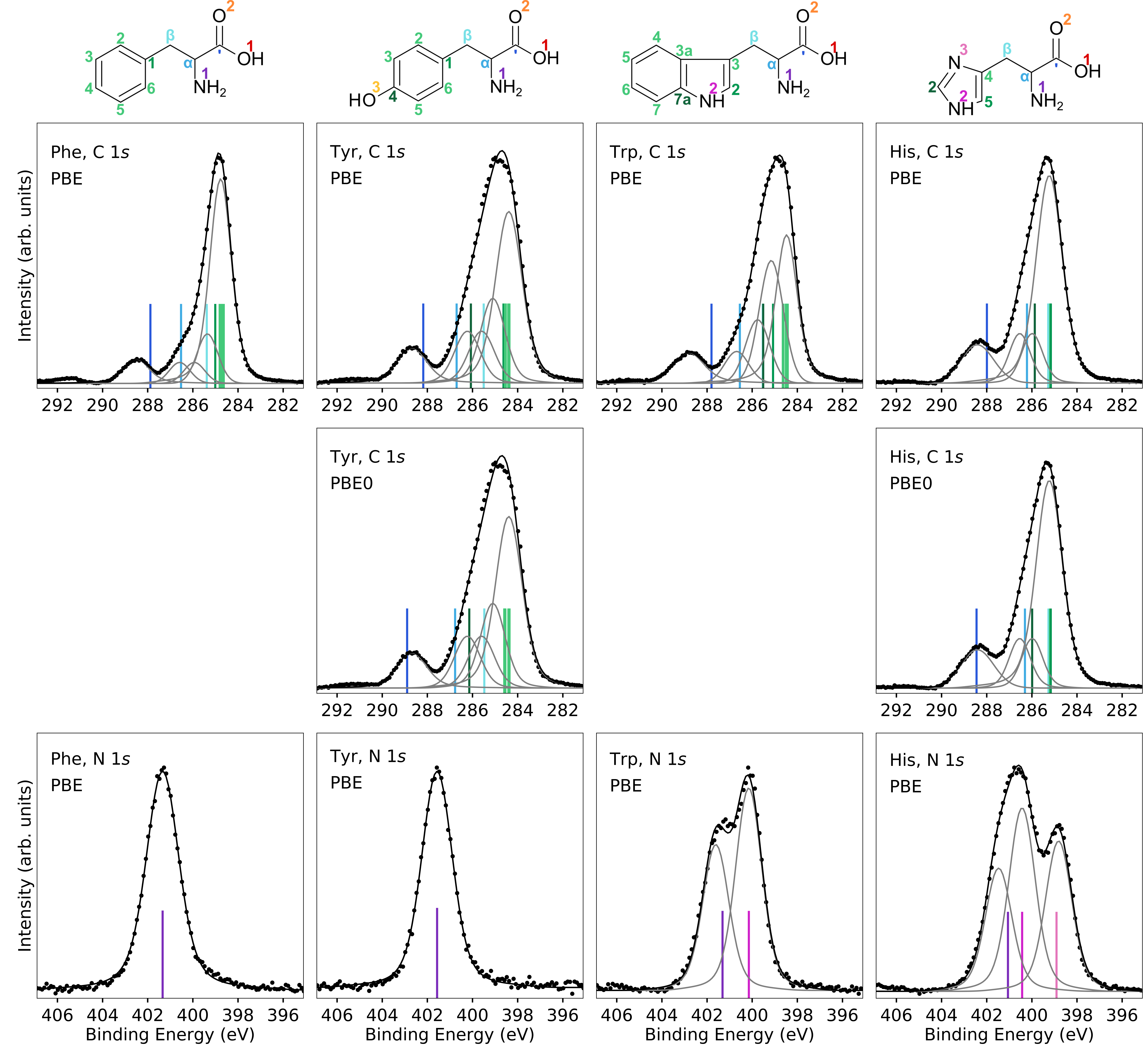}
\caption{C and N~1\textit{s} core level spectra, with experiments depicted as black dots, experimental peak fits denoted as grey/black solid lines, and calculated BEs shown as coloured vertical lines. PBE0 calculations are omitted for N~1\textit{s} due to the similarity with PBE results. Calculated BEs have been aligned with the experimental spectra by aligning with respect to the lowest BE peak, taking the average calculated BE where appropriate. A Shirley-type background has been subtracted from all core level spectra to aid comparison with theory.\label{fig:spectra}}
\end{figure*}

Where experimental core level spectra exist in the literature, they are very similar to the data presented here, albeit often with lower energy resolution~\cite{Zhang2009ElectronicSpectroscopy,Clark1976,Zubavichus2004SoftStudy,Stevens2013}. The main difference is often found in the peak fits, including the number of peaks fitted and their relative BEs and intensities. The peak fits presented here are based on robust, physically justifiable line shapes, including FWHM and L/G ratio, with the number of peaks informed from theory where needed due to overlap.
In Fig.~\ref{fig:spectra} a Shirley-type background has been subtracted to aid comparison with theory, while the relative BEs are presented in Tab.~\ref{tab:bes} alongside the calculated values. Absolute BEs resulting from the peak fits are given in the Supplementary Information. It should be noted that adventitious carbon at around 285 eV is present in all samples as is expected for XPS of ex-situ prepared powders, leading to a slight deviation from expected relative intensities.\\

\subsubsection{C~1\textit{s}}

Considering first the C~1\textit{s} BEs, it is clear that PBE0-calculated values agree more closely with experiment. The main discrepancy is that for PBE calculations the BE of C$'$ is much closer to C$^\alpha$ than for PBE0. In the worst case, Tyr, the difference between C$'$ and C$^\alpha$ is 1~eV smaller than for the experimental BEs, while the difference for PBE0-calculated BEs is much closer to experiment.  This is clearly evident in Fig.~\ref{fig:spectra}. However, as previously discussed, the relative BEs of all C atoms other than C$'$ are in very similar positions relative to each other for both PBE and PBE0. As a result, where the calculated BEs are aligned with respect to the lowest BE peak as in Fig.~\ref{fig:spectra}, the only visible difference between PBE and PBE0 is in the position of C$'$.  This is reflected in the mean absolute error (MAE) of the BEs between experiment and theory -- taking C$'$ as a reference the MAE is at 0.2~eV or less for PBE0, while in the worst case for PBE, Trp, this is much higher at 0.9~eV. If, however, the BEs are aligned with respect to the lowest BE peak, the PBE MAEs are similar to PBE0.\\

His is the only amino acid included here for which high resolution solid state spectra have previously been reported~\cite{Stevens2013}. As the work by Stevens \textit{et al.}\ includes detailed information on the peak fits and resulting peak positions, this can be directly compared with the present results. The peak assignments made in the Stevens work agree well with what the Koopmans' level of theory predicts for gas phase His, with C$'$, C$^\alpha$ and C$^\beta$ in good agreement with the present results. The main difference lies in the assignment of the subpeaks of the aromatic C atoms C$_2$, C$_4$ and C$_5$. C$_4$ and C$_5$ are assigned an intermediate BE between C$^\beta$ and C$^\alpha$ in the Stevens work, but based on the solid state $\Delta$SCF theory results presented here, it is clear that both overlap with C$^\beta$. And whilst C$_2$ is assigned the second highest BE in the previous work, it becomes clear that it actually lies below C$^\alpha$. The peak fits presented in Fig.~\ref{fig:spectra} take into account the theoretical results and a good agreement between the two is found. \\

In addition to the main photoionisation features all C~1\textit{s} spectra include $\pi-\pi^*$ shake-up satellites at 6-7 eV above the main photoionisation peak at lowest BE with relative intensities of $\leq$3\% compared to the aromatic contribution of the C~1\textit{s} core level. This is in good agreement with observations made for many conjugated systems, including early studies of Phe, Tyr and Trp by Clark \textit{et al.}\cite{Clark1976}. The calculation of satellite features is challenging and they are not included in the theoretical calculations presented here, although we note that approaches based on both DFT and time-dependent DFT have been successfully employed for large molecules~\cite{Brena2004EquivalentPhthalocyanine,Gao2008}.\\

\subsubsection{N~1\textit{s}}

In contrast to C~1\textit{s}, where a considerable difference in PBE vs.\ PBE0-calculated BE values is observed, the N~1\textit{s} BEs are not strongly affected by the functional. The calculated BEs are closer together than the experimental BEs, however the MAEs are in line with those for C~1\textit{s}. Only Trp and His have more than one N atom and therefore only these two will be discussed in detail in this section.  The BEs for the molecular subspecies series as well as gas phase amino acids are given in the Supplementary Information.\\

For Trp, a big change in the difference between the BEs for N$^1$ and N$^2$ is observed when going from Koopmans' to $\Delta$SCF for the gas phase calculations, but in both cases N$^2$ is at a higher BE, in agreement with calculations from Zhang \textit{et al.}~\cite{Zhang2009ElectronicSpectroscopy}. The order of the calculated BEs in (\textbf{25}) and (\textbf{26}) is also consistent with gas phase Trp. The calculations by Zhang \textit{et al.}\ also show a strong variation between conformers, particularly for N$^1$ which varies by up to 0.7~eV, which they attribute to differences in the nature of the internal hydrogen bonding present in a given conformer.  In the solid phase the BE order of N$^1$ and N$^2$ flips compared to the gas phase, which is attributed to the presence of the zwitterion state in the solid phase and the resulting intermolecular interactions. To understand the differences in the BEs observed for N atoms with varying protonation further, subspecies molecules (\textbf{41})-(\textbf{43}) were calculated. In the Koopmans' approach the BEs are in the order N$^3$$>$N$^2$$>$N$^1$, whilst this is reversed in $\Delta$SCF. Both the ordering and values from the $\Delta$SCF approach agree very well with gas phase measurements from Cavell and Allison~\cite{Cavell1977SiteSpectroscopy}. Therefore, the observed flipping of N$^1$ and N$^2$ is most likely not solely caused by intermolecular interactions but also originates from intrinsic final state effects. The molecules (\textbf{45}) and (\textbf{46}) once again reinforce the observed influence of aromatic systems on the BEs. In particular, the aromatic aniline (\textbf{46}) molecule is more affected by $\Delta$SCF, with a relative change of 0.3 eV compared to Koopmans'. Furthermore, there is a large difference between the BEs of (\textbf{45}) and (\textbf{46}) -- 0.7~eV for Koopmans' and 0.9~eV for $\Delta$SCF, where again the $\Delta$SCF results are in good agreement with the difference of 0.6~eV measured by Cavell and Allison.\\

The calculated N 1\textit{s} BEs for Trp agree well with the experimentally observed values. In the experimental N 1\textit{s} spectrum of Trp a higher intensity of the peak assigned to N$^2$ relative to N$^1$ is observed. This deviation from the 1:1 ratio of the two N components has been reported previously~\cite{Clark1976},
and is most likely caused by a partial deprotonation of the NH\textsubscript{3}\textsuperscript{+} group at the surface of the powder sample.\\

The N~1\textit{s} BEs of His show a similar sensitivity to a range of factors as for Trp.  Looking at the gas phase conformers, N$^2$ shows a consistently higher BE for (\textbf{31}) - (\textbf{33}) and all His conformers, for both Koopmans' and $\Delta$SCF results.  However, the ordering of N$^1$ and N$^3$ changes between different conformers.  Quantitatively, the BEs also vary significantly between Koopmans' and $\Delta$SCF, with N$^3$ typically being affected most strongly, although there do not appear to be any general trends.  This again highlights the importance of taking final state effects into account. Furthermore, the trend in energies cannot be explained purely by considering protonation states, but is likely influenced by both aromaticity and interactions between the two N atoms in the ring.  As with Trp, the solid state BEs are qualitatively different from the gas phase conformers, with N$^1$ now having the highest BE and N$^3$ having the lowest BE.  The fact that N$^1$ has the highest BE agrees with the behaviour in Trp, and as for Trp the change between gas and solid state BEs is likely due to a combination of the zwitterionic nature of the amino acid in the solid state as well as the related intermolecular interactions.\\

Two previous experimental studies have reported N~1\textit{s} spectra for His. Feyer \textit{et al.}\ show N~1\textit{s} core level spectra comparable to those reported here, but are not able to resolve N$^1$ and N$^2$ in their analysis~\cite{Feyer2008TheCu110}. Stevens \textit{et al.}\ report BE values of 398.8 eV (N$^3$), 400.4 eV (N$^2$), and 401.4 eV (N$^1$) for His, which are in good agreement with our measurements and peak assignments, and both agree well with the calculated values.\\

\subsubsection{O~1\textit{s}}

To complete the set of core states present in the aromatic amino acids, the O~1\textit{s} spectra are presented in the Supplementary Information. 
As with N~1\textit{s} BEs, the calculated values are not affected by the functional, and the MAE between theory and experiment is also in line with N~1\textit{s}.
However, overall, these spectra do not provide much additional information beyond what has been discussed based on the C and N~1\textit{s} results and only Tyr has more than one oxygen environment present in the solid state. In addition, O~1\textit{s} has an intrinsically high lifetime width and small magnitude of chemical shifts, which in combination with the presence of surface states, limits its usefulness for the study of amino acids.\\

\section{Conclusion}

This work presents the first detailed, systematic exploration of the core state energies of the four aromatic amino acids combining both high resolution XPS and state-of-the-art DFT. A $\Delta$SCF approach, which we have successfully developed and applied to simpler amino acids previously, is extended to amino acids with aromatic side chains and proves robust in predicting the core levels observed in XPS and all contributing local chemical environments. More than 20 additional molecular subspecies are calculated to aid in the discussion and interpretation of the amino acid core states and underpin the assignments made in experimental spectra. This approach provides further understanding and rationalisation of the often complicated and surprising changes in binding energies observed in the calculations for the solid state amino acids. This work substantially improves our understanding of the aromatic amino acids and gives crucial insights into their intra- and intermolecular structure. Furthermore, it reemphasises the need to combine theory with experiment in order to obtain an accurate and robust picture of the local chemistry and electronic structure and forms the basis for future work on conjugated molecular systems in general. \\

\section*{Acknowledgements}

AR acknowledges support from the Analytical Chemistry Trust Fund for her CAMS-UK Fellowship.
NKF acknowledges support from the Engineering and Physical Sciences Research Council (EP/L015277/1).
LER acknowledges support from an EPSRC Early Career Research Fellowship (EP/P033253/1) and the Thomas Young Centre under grant number TYC-101.
Calculations were performed on the Imperial College High Performance Computing Service and the ARCHER UK National Supercomputing Service.

\section*{References}

\bibliographystyle{iopart-num}
\bibliography{references}

\providecommand{\newblock}{}
\begin{thebibliography}{10}
\expandafter\ifx\csname url\endcsname\relax
  \def\url#1{{\tt #1}}\fi
\expandafter\ifx\csname urlprefix\endcsname\relax\def\urlprefix{URL }\fi
\providecommand{\eprint}[2][]{\url{#2}}

\bibitem{Hohenberg1964}
Hohenberg P and Kohn W 1964 {\em Phys. Rev.\/} {\bf 136} B864--B871

\bibitem{Kohn1965}
Kohn W and Sham L~J 1965 {\em Phys. Rev.\/} {\bf 140} A1133--A1138

\bibitem{Pi2020}
Pi J~M, Stella M, Fernando N~K, Lam A~Y, Regoutz A and Ratcliff L~E 2020 {\em
  J. Phys. Chem. Lett.\/} {\bf 11} 2256--2262

\bibitem{Thomas2007AdsorptionSurface}
Thomas A~G, Flavell W~R, Chatwin C~P, Kumarasinghe A~R, Rayner S~M, Kirkham
  P~F, Tsoutsou D, Johal T~K and Patel S 2007 {\em Surf. Sci.\/} {\bf 601}
  3828--3832

\bibitem{Feyer2008}
Feyer V, Plekan O, Richter R, Coreno M, Prince K~C and Carravetta V 2008 {\em
  J. Phys. Chem. A\/} {\bf 112} 7806--7815

\bibitem{Feyer2010AdsorptionAu111}
Feyer V, Plekan O, Tsud N, Ch{\'{a}}b V, Matol{\'{i}}n V and Prince K~C 2010
  {\em Langmuir\/} {\bf 26} 8606--8613

\bibitem{Reichert2010L-tyrosineScheme}
Reichert J, Schiffrin A, Auw{\"{a}}rter W, Weber-Bargioni A, Marschall M,
  Dell'Angela M, Cvetko D, Bavdek G, Cossaro A, Morgante A and Barth J~V 2010
  {\em ACS Nano\/} {\bf 4} 1218--1226

\bibitem{Zhang2009ElectronicSpectroscopy}
Zhang W, Carravetta V, Plekan O, Feyer V, Richter R, Coreno M and Prince K~C
  2009 {\em J. Chem. Phys.\/} {\bf 131}

\bibitem{Zubavichus2004SoftStudy}
Zubavichus Y, Zharnikov M, Shaporenko A, Fuchs O, Weinhardt L, Heske C, Umbach
  E, Denlinger J~D and Grunze M 2004 {\em J. Phys. Chem. A\/} {\bf 108}
  4557--4565

\bibitem{Cardenas2006TheSpectroscopy}
Cardenas J~F and Gr{\"{o}}bner G 2006 {\em J. Electron. Spectros. Relat.
  Phenomena\/} {\bf 152} 87--90

\bibitem{Stevens2013}
Stevens J~S, De~Luca A~C, Pelendritis M, Terenghi G, Downes S and Schroeder S~L
  2013 {\em Surf. Interface Anal.\/} {\bf 45} 1238--1246

\bibitem{Boese1997CarbonPeptides}
Boese J, Osanna A, Jacobsen C and Kirz J 1997 {\em J. Electron. Spectros.
  Relat. Phenomena\/} {\bf 85} 9--15

\bibitem{Cooper2004InnerPhenylalanine}
Cooper G, Gordon M, Tulumello D, Turci C, Kaznatcheev K and Hitchcock A~P 2004
  {\em J. Electron. Spectros. Relat. Phenomena\/} {\bf 137-140} 795--799

\bibitem{Wang2008}
Wang F 2008 {\em J. Phys.: Conf. Ser.\/} {\bf 141} 012019

\bibitem{Ganesan2009}
Ganesan A and Wang F 2009 {\em J. Chem. Phys.\/} {\bf 131}

\bibitem{Ganesan2014}
Ganesan A, Mohammadi N and Wang F 2014 {\em RSC Adv.\/} {\bf 4} 8617

\bibitem{Wang2014}
Wang F and Ganesan A 2014 {\em RSC Adv.\/} {\bf 4} 60597--60608

\bibitem{Destro1988}
Destro R, Marsh R~E and Bianchi R 1988 {\em J. Phys. Chem\/} {\bf 92} 966--973

\bibitem{Ihlefeldt2014}
Ihlefeldt F~S, Pettersen F~B, Von~Bonin A, Zawadzka M and G{\"{o}}rbitz C~H
  2014 {\em Angew. Chem. Int.\/} {\bf 53} 13600--13604

\bibitem{Mostad1972}
{Mostad A}, {Nissen H M} and {Romming C} 1972 {\em Acta Chem. Scand.\/} {\bf
  26} 3819--3833

\bibitem{Gorbitz2015}
G{\"{o}}rbitz C~H, T{\"{o}}rnroos K~W and Day G~M 2012 {\em Acta Crystallogr.
  B\/} {\bf 68} 549--557

\bibitem{Fronczek2016}
Fronczek F~R 2016 {\em CSD Communication\/}  CCDC 1446832

\bibitem{Purushotham2012}
Purushotham U, Vijay D and Narahari~Sastry G 2012 {\em J. Comp. Chem.\/} {\bf
  33} 44--59

\bibitem{Ropo2016}
Ropo M, Schneider M, Baldauf C and Blum V 2016 {\em Sci. Data\/} {\bf 3} 160009

\bibitem{Kahk2019}
Kahk J~M and Lischner J 2019 {\em Phys. Rev. Mater.\/} {\bf 3}

\bibitem{Ozaki2017}
Ozaki T and Lee C~C 2017 {\em Phys. Rev. Lett.\/} {\bf 118} 026401

\bibitem{Genovese2008}
Genovese L, Neelov A, Goedecker S, Deutsch T, Ghasemi S~A, Willand A, Caliste
  D, Zilberberg O, Rayson M, Bergman A and Schneider R 2008 {\em J. Chem.
  Phys.\/} {\bf 129} 014109

\bibitem{Ratcliff2020}
Ratcliff L~E, Dawson W, Fisicaro G, Caliste D, Mohr S, Degomme A, Videau B,
  Cristiglio V, Stella M, D’Alessandro M, Goedecker S, Nakajima T, Deutsch T
  and Genovese L 2020 {\em J. Chem. Phys.\/} {\bf 152} 194110

\bibitem{Goedecker1996}
Goedecker S, Teter M and Hutter J 1996 {\em Phys. Rev. B\/} {\bf 54} 1703--1710

\bibitem{Hartwigsen1998}
Hartwigsen C, Goedecker S and Hutter J 1998 {\em Phys. Rev. B\/} {\bf 58}
  3641--3662

\bibitem{Harrison2016}
Harrison R~J, Beylkin G, Bischoff F~A, Calvin J~A, Fann G~I, Fosso-Tande J,
  Galindo D, Hammond J~R, Hartman-Baker R, Hill J~C, Jia J, Kottmann J~S, Ou
  M~J, Pei J, Ratcliff L~E, Reuter M~G, Richie-Halford A~C, Romero N~A, Sekino
  H, Shelton W~A, Sundahl B~E, Thornton W~S, Valeev E~F,
  V{\'{a}}zquez-Mayagoitia {\'{A}}, Vence N, Yanai T and Yokoi Y 2016 {\em SIAM
  J. Sci. Comput.\/} {\bf 38} S123--S142

\bibitem{Ratcliff2019}
Ratcliff L~E, Thornton W~S, V{\'{a}}zquez-Mayagoitia {\'{A}} and Romero N~A
  2019 {\em J. Phys. Chem. A\/} {\bf 123} 4465--4474

\bibitem{Anderson2019}
Anderson J, Harrison R~J, Sekino H, Sundahl B, Beylkin G, Fann G~I, Jensen S~R
  and Sagert I 2019 {\em J. Comput. Physics: X\/} {\bf 4} 100033

\bibitem{Clark2005}
Clark S~J, Segall M~D, Pickard C~J, Hasnip P~J, Probert M~I, Refson K and Payne
  M~C 2005 {\em Z. Kristall.\/} {\bf 220} 567--570

\bibitem{Monkhorst1976}
Monkhorst H~J and Pack J~D 1976 {\em Phys. Rev. B\/} {\bf 13} 5188--5192

\bibitem{Grimme2006}
Grimme S 2006 {\em J. Comput. Chem.\/} {\bf 27} 1787--1799

\bibitem{Perdew1996}
Perdew J~P, Burke K and Ernzerhof M 1996 {\em Phys. Rev. Lett.\/} {\bf 77}
  3865--3868

\bibitem{Adamo1999}
Adamo C and Barone V 1999 {\em J. Chem. Phys.\/} {\bf 110} 6158--6170

\bibitem{Besley2009}
Besley N~A, Gilbert A~T and Gill P~M 2009 {\em J. Chem. Phys.\/} {\bf 130}
  124308

\bibitem{Momma2008}
Momma K and Izumi F 2008 {\em J. Appl. Crystallogr.\/} {\bf 41} 653--658

\bibitem{Myrseth2007}
Myrseth V, B{\o}rve K~J and Thomas T~D 2007 {\em J. Org. Chem.\/} {\bf 72}
  5715--5723

\bibitem{Ohta1975Core-electronSpectroscopy}
Ohta T, Fujikawa T and Kuroda H 1975 {\em Bull. Chem. Soc. Jpn.\/} {\bf 48}
  2017--2024

\bibitem{Clark1976}
Clark D~T, Peeling J and Colling L 1976 {\em Biochim. Biophys. Acta\/} {\bf
  453} 533--545

\bibitem{Brena2004EquivalentPhthalocyanine}
Brena B, Luo Y, Nyberg M, Carniato S, Nilson K, Alfredsson Y, {\AA}hlund J,
  M{\aa}rtensson N, Siegbahn H and Puglia C 2004 {\em Phys. Rev. B\/} {\bf 70}
  1--6

\bibitem{Gao2008}
Gao B, Wu Z and Luo Y 2008 {\em J. Chem. Phys.\/} {\bf 128} 234704

\bibitem{Cavell1977SiteSpectroscopy}
Cavell R~G and Allison D~A 1977 {\em J. Am. Chem. Soc.\/} {\bf 99} 4203--4204

\bibitem{Feyer2008TheCu110}
Feyer V, Plekan O, Sk{\'{a}}la T, Ch{\'{a}}b V, Matol{\'{i}}n V and Prince K~C
  2008 {\em J. Phys. Chem. B\/} {\bf 112} 13655--13660

\end{thebibliography}

\end{document}


\author{Anna Regoutz}         \affiliation{\UCL}
\author{Marta S.\ Wolinska}     \affiliation{\ICL}
\author{Nathalie K.\ Fernando}     \affiliation{\UCL}
\author{Laura E.\ Ratcliff}     \affiliation{\ICL}

\date{\today}

\title{\LARGE Supplementary Information\\
\vspace{5pt}
\Large A Combined Density Functional Theory and X-ray Photoelectron Spectroscopy Study of the Aromatic Amino Acids}
\maketitle

\begin{table*}[h]
\centering
\begin{threeparttable}
\caption{\label{tab:crystals} Experimental (`Exp') and theoretical (`Theor') unit cell parameters for the PBE-relaxed crystals.  Lattice parameters, angles, and differences (`Diff') are in \AA,  \textdegree\, and \% of the experimental value, respectively.}
\begin{tabular*} {1.0\textwidth}{l @{\extracolsep{\fill}} rrrrrr}
\hline\hline
& $a$ & $b$ & $c$ & $\alpha$ & $\beta$ & $\gamma$ \\
\cline{1-1} \cline{2-4} \cline{5-7}\\[-2.5ex]
\textbf{Ala} \\
Exp\tnote{a} & 5.93	& 12.26	& 5.79	& 90.00 &	90.00 &	90.00\\
Theor & 5.96	& 12.23 &	5.86 &	90.00	& 90.00	& 90.00\\
Diff & 0.6	& -0.3	& 1.2	& 0.0	& 0.0	& 0.0\\
\cline{1-1} \cline{2-4} \cline{5-7}\\[-2.5ex]
\textbf{Phe}\\
Exp\tnote{b} & 8.78	& 6.00	& 31.02	& 90.00	& 96.92	& 90.00\\
Theor & 8.82	& 5.99	& 30.81	& 90.00	& 96.98	& 90.00\\
Diff & 0.4	& -0.1	& -0.7	& 0.0	& 0.1	& 0.0\\
\cline{1-1} \cline{2-4} \cline{5-7}\\[-2.5ex]
\textbf{Tyr}\\
Exp\tnote{c} & 6.91	& 21.12	& 5.83	& 90.00	& 90.00	& 90.00\\
Theor & 6.70	& 21.29	& 5.87	& 90.00	& 90.00	& 90.00\\
Diff & -3.0 & 0.8	& 0.7	& 0.0	& 0.0	& 0.0\\
\cline{1-1} \cline{2-4} \cline{5-7}\\[-2.5ex]
\textbf{Trp} \\
Exp\tnote{d} & 11.43 &	11.46	& 35.61	& 84.42	& 87.69	& 60.10\\
Theor & 11.43	& 11.54	& 35.51	& 84.32	& 87.70	& 60.33\\
Diff & 0.0	& 0.7	& -0.3	& -0.1	& 0.0	& 0.4\\
\cline{1-1} \cline{2-4} \cline{5-7}\\[-2.5ex]
\textbf{His}\\
Exp\tnote{e} & 5.16	& 7.30	& 9.47	& 90.00	& 97.58	& 90.00\\
Theor & 5.17	& 7.28	& 9.53	& 90.00	& 96.45	& 90.00\\
Diff & 0.2	& -0.3	& 0.5	& 0.0	& -1.2	& 0.0\\
\hline\hline
\end{tabular*}
\begin{tablenotes}
 \item[a] Ref.~\onlinecite{Destro1988}.  
 \item[b] Ref.~\onlinecite{Ihlefeldt2014}.
 \item[c] Ref.~\onlinecite{Mostad1972}.
 \item[d] Ref.~\onlinecite{Gorbitz2015}.
 \item[e] Ref.~\onlinecite{Fronczek2016}.
 \end{tablenotes}
\end{threeparttable}
\end{table*}

\begin{figure*}[h]
\centering

\subfigure[Ala] {\includegraphics[width=0.2\textwidth]{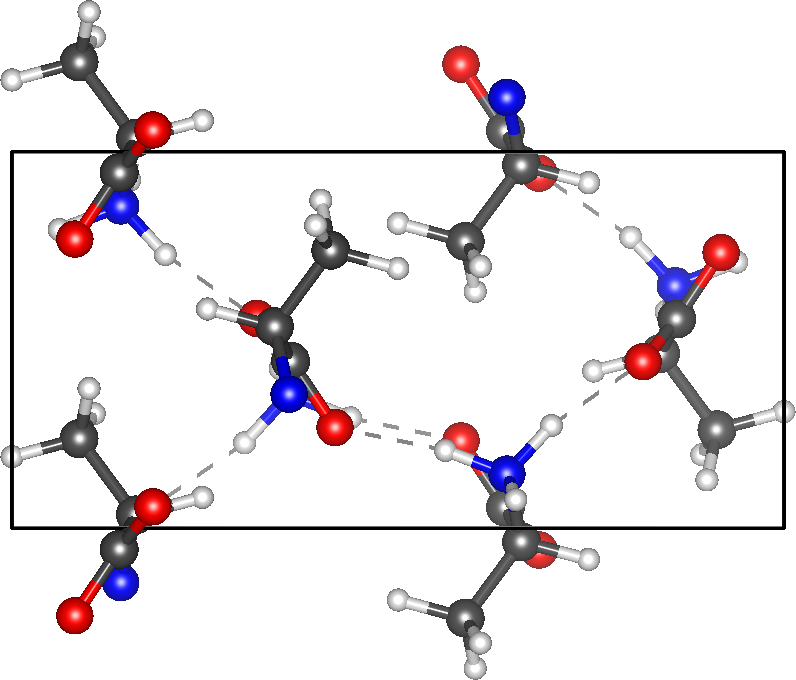}\label{fig:ala_crystal}}
\hspace{4pt}
\subfigure[Phe] {\includegraphics[width=0.45\textwidth]{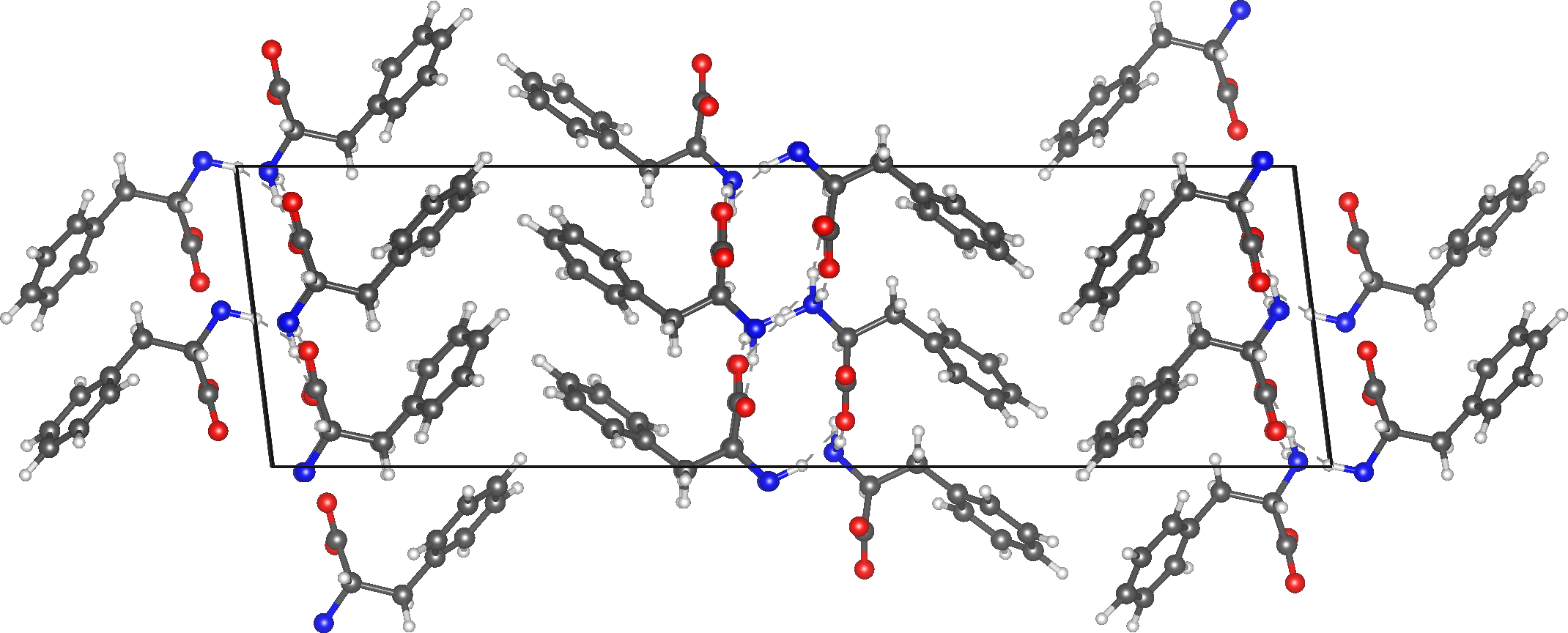}\label{fig:phe_crystal}}
\hspace{4pt}
\subfigure[His] {\includegraphics[width=0.25\textwidth]{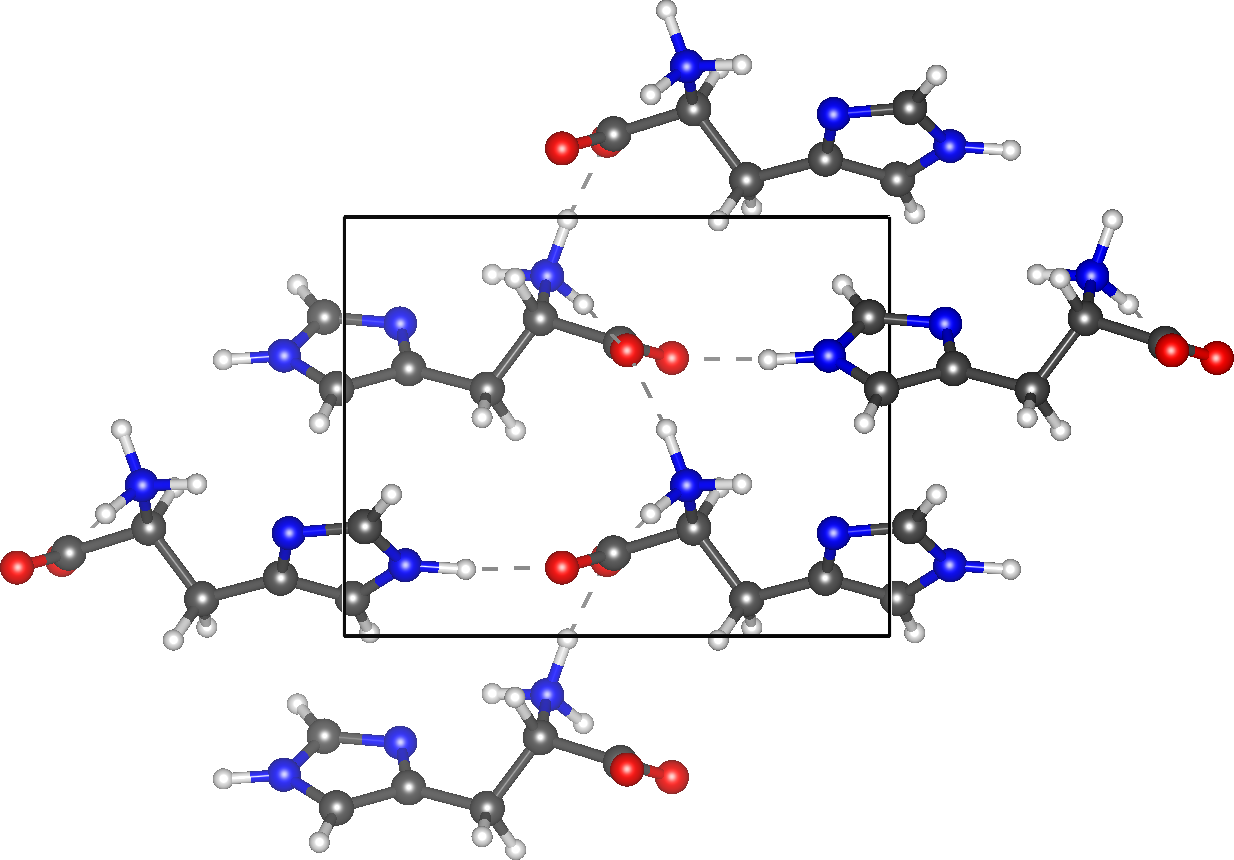}\label{fig:his_crystal}}

\subfigure[Tyr] {\includegraphics[width=0.45\textwidth]{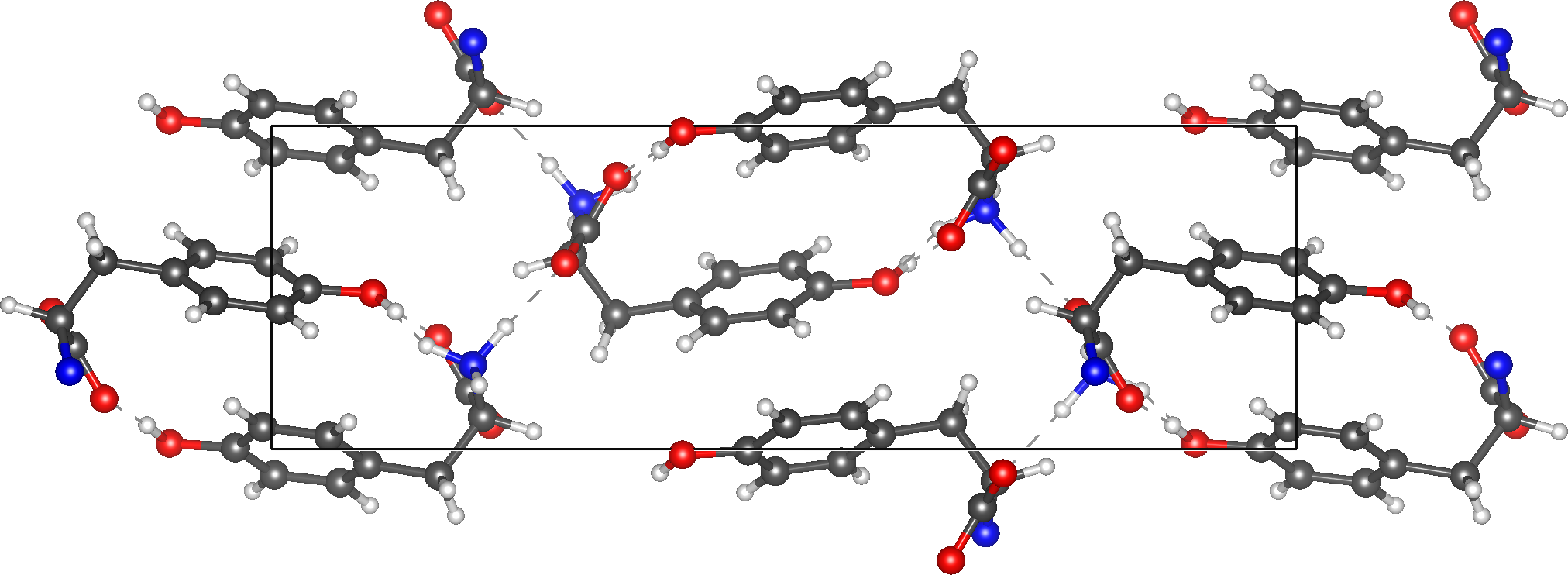}\label{fig:tyr_crystal}}
\hspace{8pt}
\subfigure[Trp] {\includegraphics[width=0.45\textwidth]{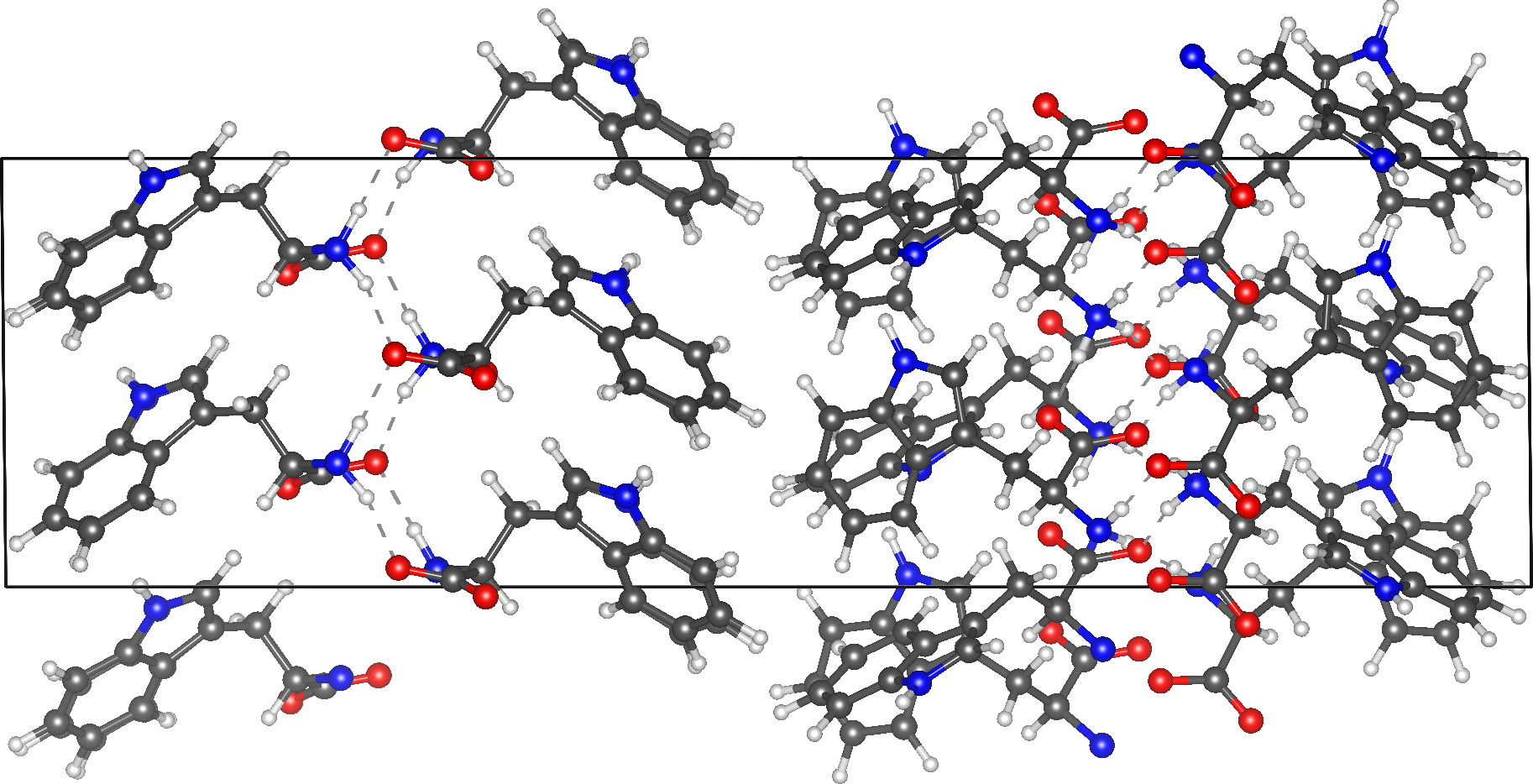}\label{fig:trp_crystal}}
\caption{Relaxed crystal structures of (a) Ala, (b) Phe, (c) His, (d) Tyr, and (e) Trp, where C/O/N/H atoms are represented in grey/red/blue/white. The black lines indicate the cell boundaries. 
\label{fig:structures}}
\end{figure*}

\begin{figure*}[h]
\centering
\subfigure[$\rho_\mathbf{2} - \rho_\mathbf{1}$] {\includegraphics[scale=0.14]{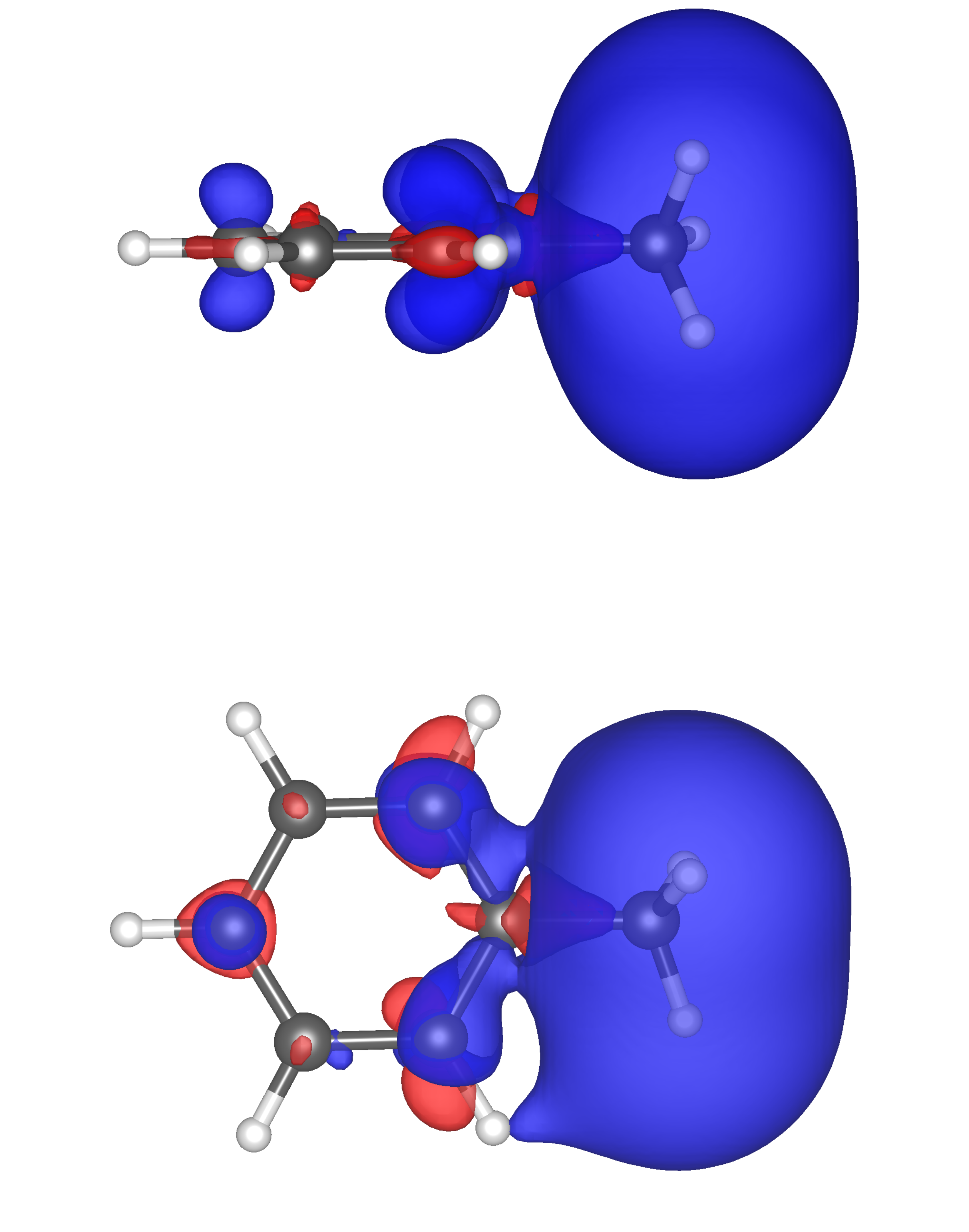}\label{fig:dens_2_1}}
\hspace{8pt}
\subfigure[$\rho_\mathbf{3} - \rho_\mathbf{2}$] {\includegraphics[scale=0.14]{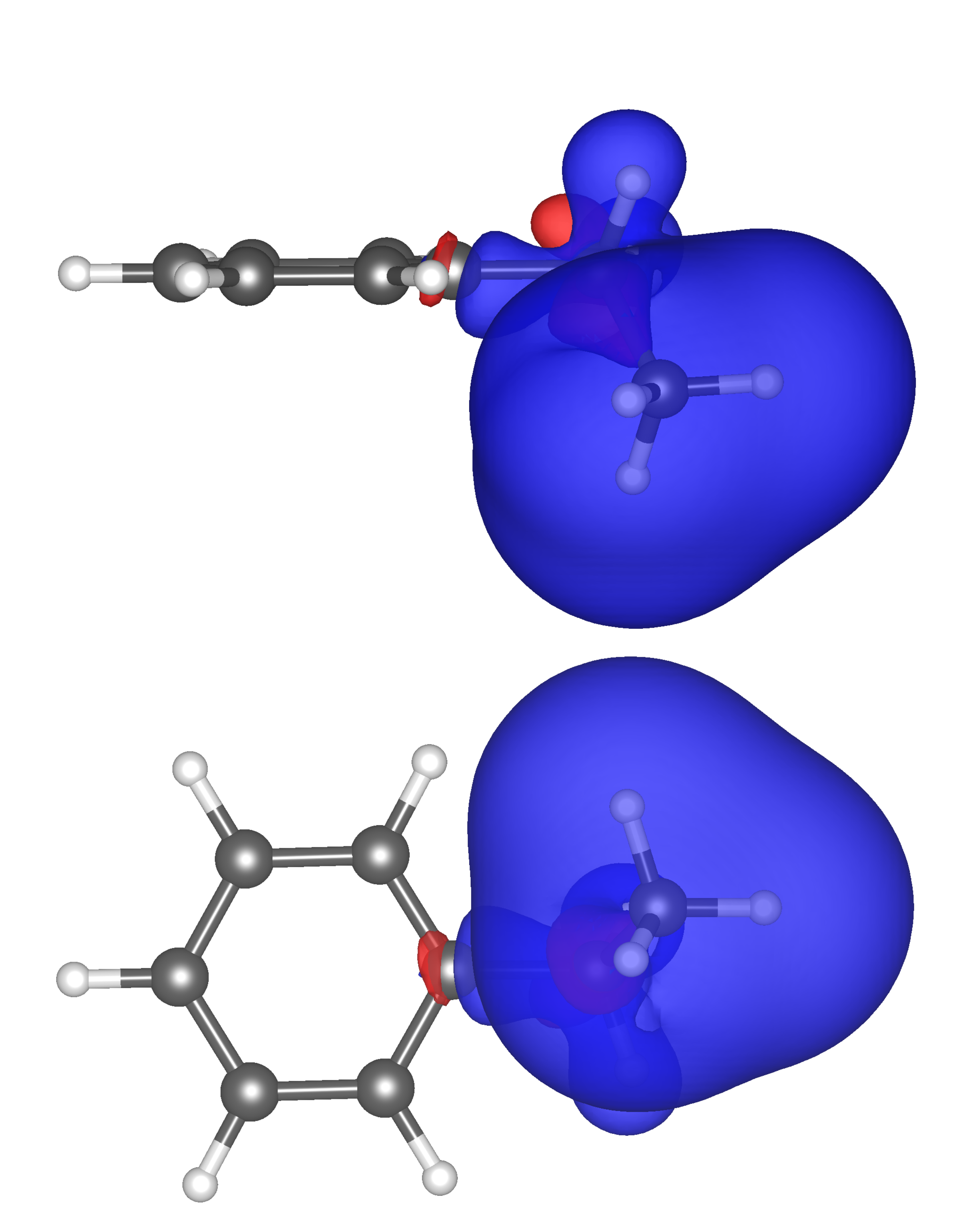}\label{fig:dens_3_2}}
\hspace{8pt}
\subfigure[$\rho_\mathbf{12} - \rho_\mathbf{11}$] {\includegraphics[scale=0.14]{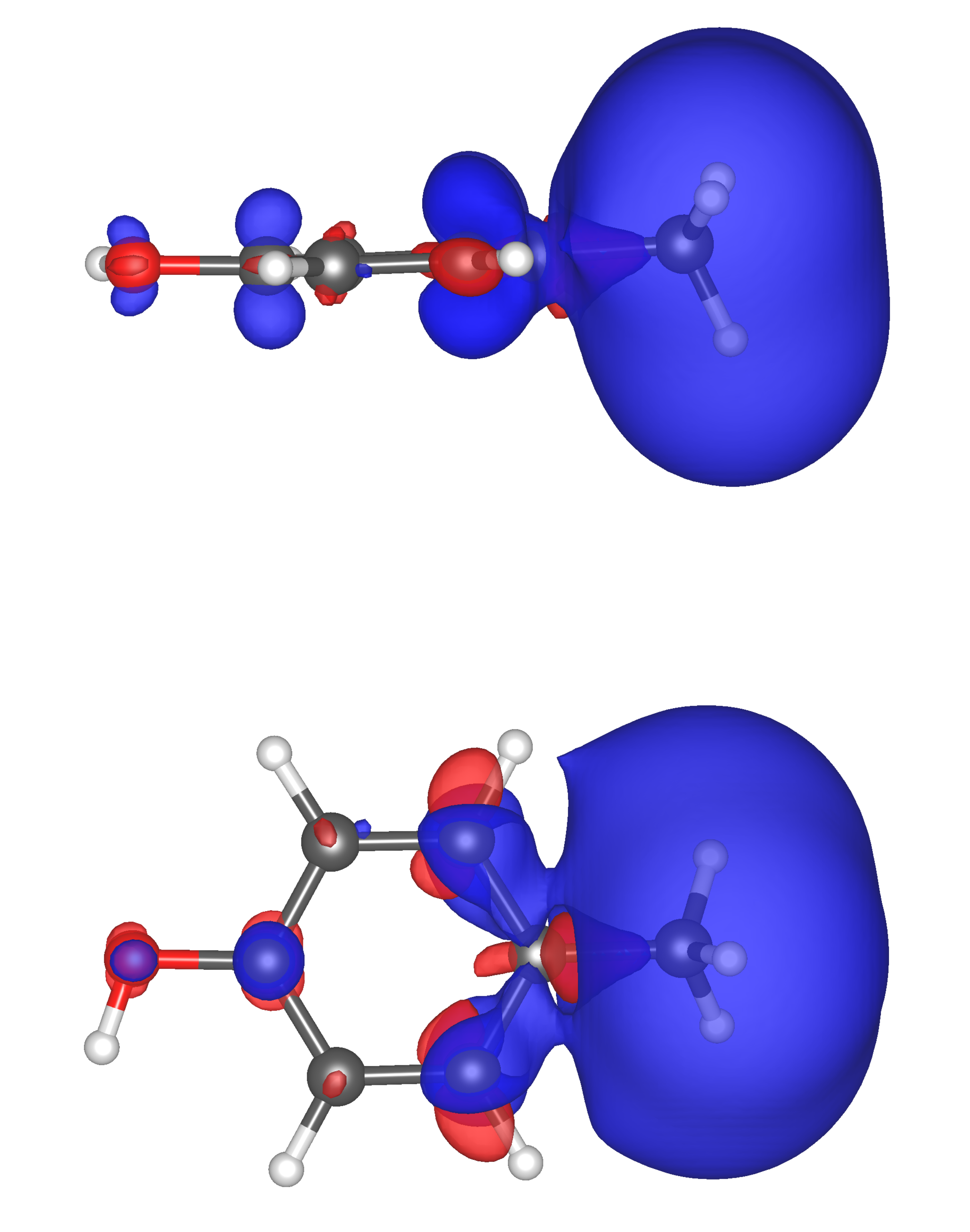}\label{fig:dens_12_11}}
\hspace{8pt}
\subfigure[$\rho_\mathbf{13} - \rho_\mathbf{12}$] {\includegraphics[scale=0.14]{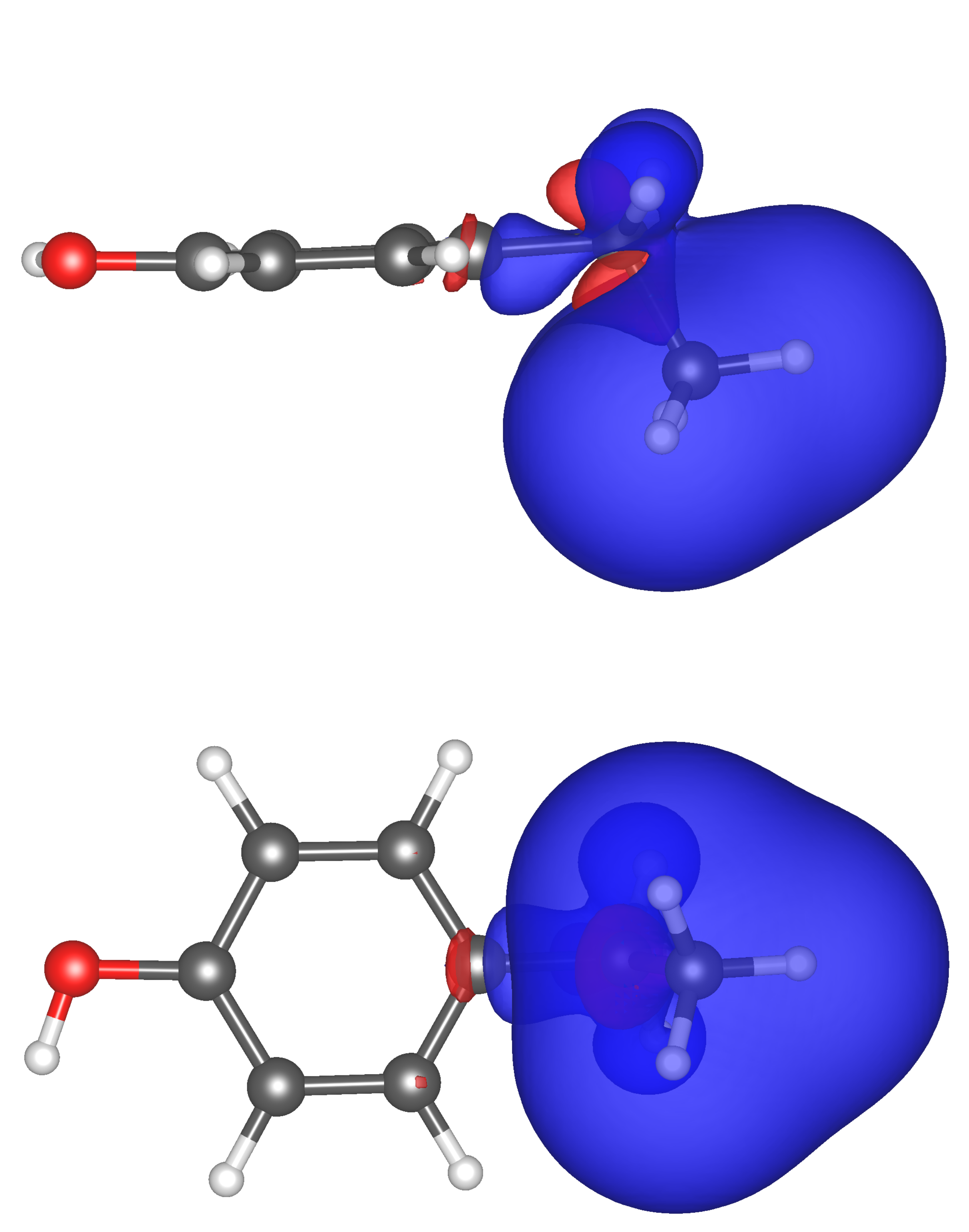}\label{fig:dens_13_12}}
\caption{Difference between calculated densities of pairs of molecules, where two views are shown for each pair of molecules and blue indicates an increase in density and red a decrease. Densities were calculated using PBE in BigDFT, with each molecule placed in a cubic supercell with a side length of 25~\AA\ for convenience. All atoms were held fixed between the series of molecules, except for the most recently added functional group. All other computational parameters are as described in the article text. 
\label{fig:densities}}
\end{figure*}

\clearpage

\begin{figure*}[ht]
\centering
\includegraphics[width=1.0\textwidth]{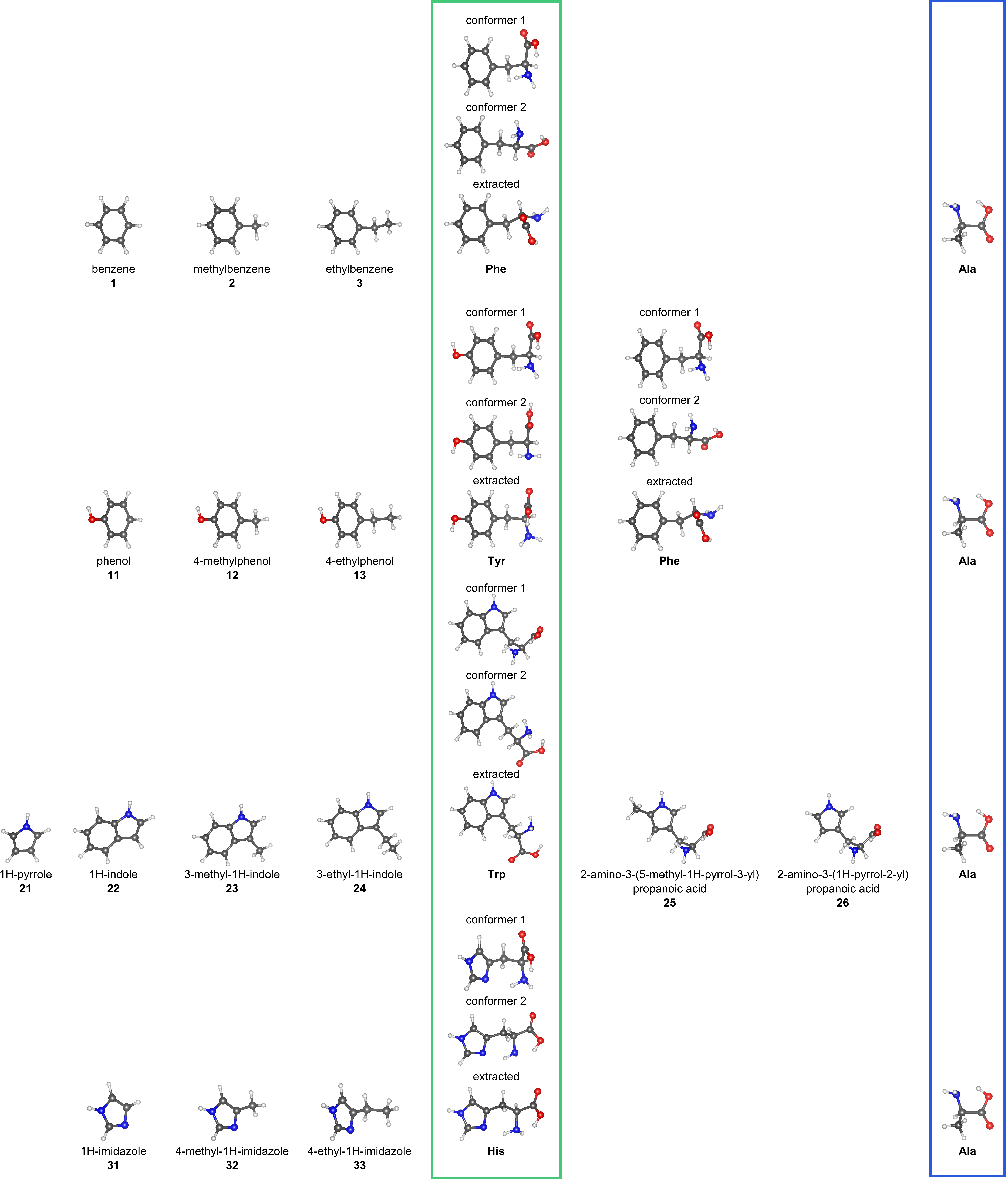}
\caption{Relaxed atomic structures of the conformers of the molecular subspecies, and both the conformers and structures extracted from the relaxed crystals for the gas phase amino acids. C/O/N/H atoms are depicted in grey/red/blue/white. \label{fig:conformers}}
\end{figure*}

\clearpage

\begin{figure}[h]
\centering
\includegraphics[scale=0.5]{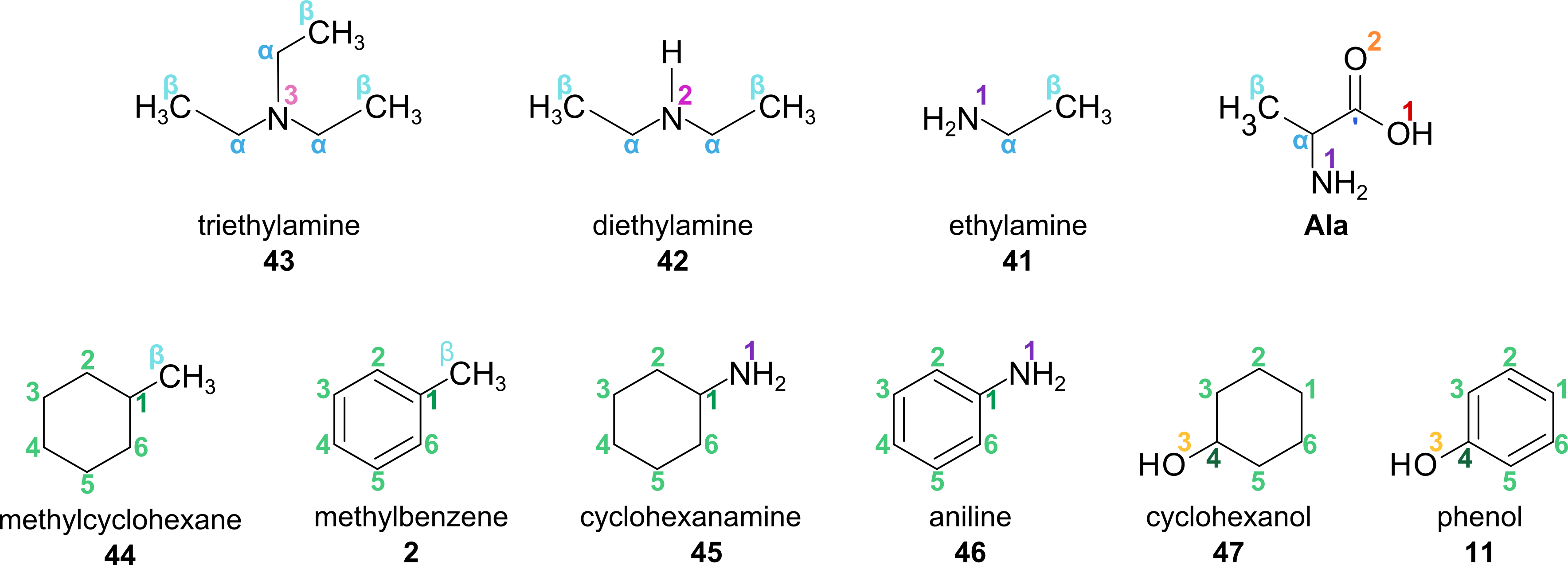}
\caption{Molecular structure of the additional calculated molecular subspecies as well as Ala and molecules (\textbf{2}) and (\textbf{11}) of the main set, including atom labels used throughout this work. The molecules included are ethylamine (\textbf{41}), diethylamine (\textbf{42}), triethylamine (\textbf{43}), methylcyclohexane (\textbf{44}), methylbenzene (\textbf{2}), cyclohexanamine (\textbf{45}), aniline (\textbf{46}), cyclohexanol (\textbf{47}), and phenol (\textbf{11}). \label{fig:extra_mols_schem}}
\end{figure}

\begin{figure*}[ht]
\centering
\includegraphics[scale=0.5]{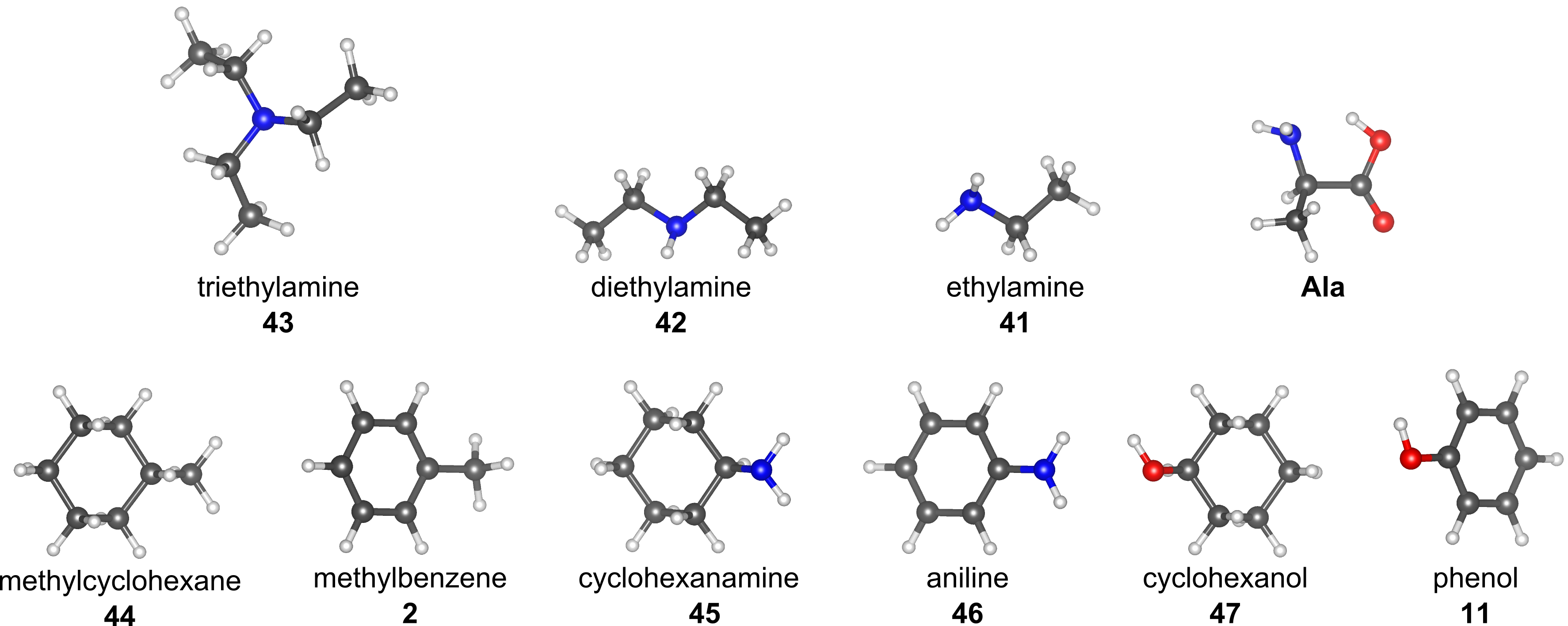}
\caption{Relaxed atomic structures of the conformers of the additional calculated molecular subspecies and gas phase Ala. C/O/N/H atoms are depicted in grey/red/blue/white. \label{fig:extra_conformers}}
\end{figure*}

\begin{figure}[h]
\centering
\includegraphics[scale=0.25]{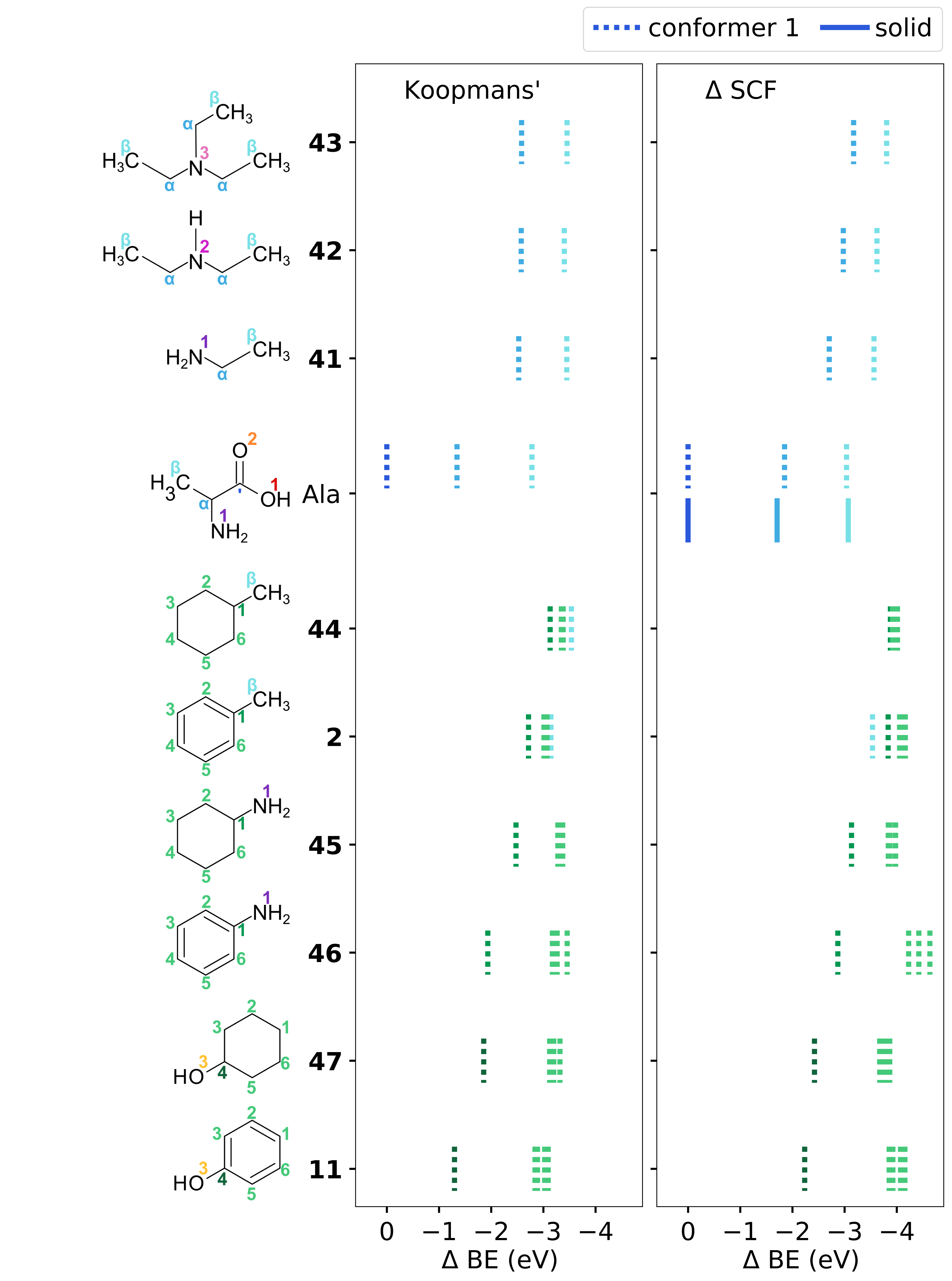}
\caption{PBE-calculated C~1\textit{s} BEs for the additional calculated molecular subspecies and Ala. BEs are relative to Ala C$'$ in gas phase (solid state) for gas phase conformer (Ala solid state) calculations.\label{fig:extra_mols}}
\end{figure}

\begin{table*}[h]
\centering
\begin{threeparttable}
\caption{\label{tab:c_phe_series_bes} C~1\textit{s} PBE-calculated BEs for the Phe series of gas phase amino acid conformers, the extracted Phe molecule, and the subspecies molecules, as well as the solid state calculations. Gas phase BEs are relative to Ala C$'$, while solid state BEs are relative to C$'$ of that amino acid.}
\begin{tabular*} {1.0\textwidth}{l @{\extracolsep{\fill}} rrrrrrrrr}
\hline\hline

 &	\textbf{1}	& \textbf{2} & 	\textbf{3}	& \multicolumn{4}{c}{\textbf{Phe}} & \multicolumn{2}{c}{\textbf{Ala}}\\
& conf 1	& conf 1	& conf 1	& conf 1	& conf 2	& extract	& solid & conf 1 & solid\\
\cline{1-1} \cline{2-2} \cline{3-3} \cline{4-4} \cline{5-8} \cline{9-10} \\[-2.5ex]
\textbf{Koopmans'}\\
C$'$ &	- & -  & - & 	0.1	& 0.1	& 0.1	& - & 0.0 & - \\
C$^\alpha$ 	& -  & -  & -3.3	& -1.3	& -1.3	& -1.7	& - & -1.3 & -\\
C$^\beta$ 	& - &	-3.1 &	-3.0 &	-2.5 &	-2.5 &	-2.5 &	- & -2.8 & -\\
C$_1$ &	-2.9 &	-2.7 &	-2.8 &	-2.4 &	-2.3 &	-2.7	& - &  - & -\\
C$_2$ &	-2.9 &	-3.1 &	-3.1 &	-2.8 &	-2.6 &	-3.1	& - & - & -\\
C$_3$ &	-2.9 &	-3.0 &	-3.0 &	-2.8 &	-2.6 &	-3.1	& - & - & -\\
C$_4$ &	-2.9 &	-3.1 &	-3.1 &	-2.8 &	-2.7 &	-3.1	& - & - & -\\
C$_5$ &	-2.9 &	-3.0 &	-3.0 &	-2.7 &	-2.6 &	-3.1	& - & - & -\\
C$_6$ &	-2.9 &	-3.1 &	-3.1 &	-2.7 &	-2.6 &	-3.1	& - & - & -\\
\cline{1-1} \cline{2-2} \cline{3-3} \cline{4-4} \cline{5-8} \cline{9-10} \\[-2.5ex]
\textbf{$\Delta$SCF}\\
C$'$ &	- & -  & - &	-0.1 &	-0.1	& -0.1	& 0.0	& 0.0 & 0.0\\
C$^\alpha$ 	& -  & -  & -3.7 &	-2.2 &	-2.2 &	-2.5	 & -1.4	& -1.8 & -1.7\\
C$^\beta$ 	& - &	-3.5 &	-3.6 &	-3.3 &	-3.3 &	-3.3 &	-2.5 &	-3.0 & -3.1\\
C$_1$ &	-3.9 &	-3.8 &	-4.0 &	-3.7 &	-3.6 &	-3.9 &	-2.9	& - & -\\
C$_2$ &		-3.9 &	-4.2 &	-4.2 &	-4.0 &	-3.8 &	-4.2 &	-3.2	& - & -\\
C$_3$ &		-3.9	& -4.1	& -4.1	& -3.9	& -3.8	& -4.1	& -3.1	& - & -\\
C$_4$ &	-3.9 &	-4.1 &	-4.2 &	-3.9 &	-3.8	& -4.2	& -3.2	& - & -\\
C$_5$ &		-3.9 &	-4.1 &	-4.1 &	-3.8 &	-3.7	& -4.1	& -3.1	& - & -\\
C$_6$ &	-3.9 &	-4.2 &	-4.2 &	-3.9 &	-3.8 &	-4.2 &	-3.2	& - & -\\
\hline\hline
\end{tabular*}
\end{threeparttable}
\end{table*}

\begin{table*}[h]
\centering
\begin{threeparttable}
\caption{\label{tab:c_tyr_series_bes} C~1\textit{s} PBE-calculated BEs for the Tyr series of gas phase amino acid conformers, the extracted Tyr molecule, and the subspecies molecules, as well as the solid state calculations. Gas phase BEs are relative to Ala C$'$, while solid state BEs are relative to C$'$ of that amino acid.}
\begin{tabular*} {1.0\textwidth}{l @{\extracolsep{\fill}} rrrrrrrrrrrrr}
\hline\hline

 &	\textbf{11}	& \textbf{12} & 	\textbf{13}	& \multicolumn{4}{c}{\textbf{Tyr}} & \multicolumn{4}{c}{\textbf{Phe}} &\multicolumn{2}{c}{\textbf{Ala}}\\
& conf 1	& conf 1	& conf 1	& conf 1	& conf 2	& extract	& solid 	& conf 1	& conf 2	& extract	& solid & conf 1 & solid\\
\cline{1-1} \cline{2-2} \cline{3-3} \cline{4-4} \cline{5-8}  \cline{9-12} \cline{13-14} \\[-2.5ex]
\textbf{Koopmans'}\\
C$'$ & - & - & - & 	0.0	& 0.3	& -0.1	& - & 0.1	& 0.1	& 0.1	& - & 0.0 & -\\
C$^\alpha$ 	& - & - & -3.4	& -1.4	& -1.6	& -1.7	& - & -1.3	 &  -1.3	& -1.7	& - &  -1.3 & -\\
C$^\beta$ 	& - & -3.2	& -3.0	& -2.6	& -2.7	& -2.7	& - & -2.5	& -2.5 &	-2.5 & - & 	-2.8 & -\\
C$_1$ &	-3.1 &	-2.9 &	-3.0	& -2.6	& -2.8	& -2.9 & - & 	-2.4 &	-2.3 &	-2.7 & - & - & - \\
C$_2$ & -2.9 &	-3.0 &	-3.0	& -2.7	& -2.9	& -3.0	& - & -2.8	& -2.6	 & -3.1 & - & -  & -\\
C$_3$ & -3.0 &	-2.9 &	-3.0 &	-2.7 &	-3.0 &	-3.1 & - & 	-2.8 &	-2.6	 &-3.1 & - & - & -	\\
C$_4$ & -1.3 &	-1.4 &	-1.4 &	-1.1 &	-1.3 &	-1.4 &	- & -2.8 &	-2.7 &	-3.1 & - & - & -\\
C$_5$ &	-2.9 &	-3.1 &	-3.1 &	-2.8	& -2.8	& -2.9	& - & -2.7 &	-2.6 &	-3.1 & - & -  & -\\
C$_6$ &	-2.8 & 	-3.0 &	-3.0 &	-2.6 &	-2.8 &	-2.8 &	- & -2.7 &	-2.6	& -3.1 & - & - & - \\
\cline{1-1} \cline{2-2} \cline{3-3} \cline{4-4} \cline{5-8}  \cline{9-12} \cline{13-14} \\[-2.5ex]
\textbf{$\Delta$SCF}\\
C$'$ & - & - & - & -0.2 &	-0.2 &	-0.5	& 0.0	& -0.1	& -0.1	& -0.1	& 0.0	& 0.0 & 0.0\\
C$^\alpha$ 	& - & - & -3.8	& -2.3	& -2.5	& -2.5	& -1.5	& -2.2	& -2.2	& -2.5	& -1.4	& -1.8 & -1.7\\
C$^\beta$ 	& - & -3.6	& -3.7	& -3.4	& -3.5	& -3.4	& -2.7	& -3.3	& -3.3	& -3.3	& -2.5	& -3.0 & -3.1\\
C$_1$ & -4.2	& -4.1	& -4.2	& -3.9	& -4.1	& -4.1	& -3.6	& -3.7	& -3.6	& -3.9	& -2.9 & - & -\\
C$_2$ &	-3.9 &	-4.1	& -4.2	& -4.0	& -4.1	& -4.2	& -3.7	& -4.0	& -3.8	& -4.2	& -3.2 & -  & -\\
C$_3$ &	-4.1	& -4.1	& -4.1	& -3.9	& -4.2	& -4.2	& -3.8	& -3.9	& -3.8	& -4.1	& -3.1 & - & -\\
C$_4$ &	-2.2	& -2.5	& -2.5	& -2.2	& -2.4	& -2.5	& -2.1	& -3.9	& -3.8	& -4.2	& -3.2 & - & -\\
C$_5$ &	-3.9 &	-4.2 &	-4.3 &	-4.0 &	-4.0 &	-4.0 &	-3.8 &	-3.8 &	-3.7 &	-4.1 &	-3.1 & - & -\\
C$_6$ &	-3.9& 	-4.2	& -4.2	& -3.8	& -4.1	& -4.0	& -3.6	& -3.9	& -3.8	& -4.2	& -3.2	& - & -\\
\hline\hline
\end{tabular*}
\end{threeparttable}
\end{table*}

\begin{table*}[h]
\centering
\begin{threeparttable}
\caption{\label{tab:c_trp_series_bes} C~1\textit{s} PBE-calculated BEs for the Trp series of gas phase amino acid conformers, the extracted Trp molecule, and the subspecies molecules, as well as the solid state calculations. Gas phase BEs are relative to Ala C$'$, while solid state BEs are relative to C$'$ of that amino acid.}
\begin{tabular*} {1.0\textwidth}{l @{\extracolsep{\fill}} rrrrrrrrrrrr}
\hline\hline

 &	\textbf{21}	& \textbf{22} & 	\textbf{23} & 	\textbf{24} & \multicolumn{4}{c}{\textbf{Trp}} &  	\textbf{25} & 	\textbf{26}& \multicolumn{2}{c}{\textbf{Ala}}\\
& conf 1	& conf 1	& conf 1 	& conf 1	& conf 1	& conf 2	& extract	& solid 	& conf 1 	& conf 1 & conf 1 & solid\\
\cline{1-1} \cline{2-2} \cline{3-3} \cline{4-4} \cline{5-5}  \cline{6-9} \cline{10-10} \cline{11-11} \cline{12-13} \\[-2.5ex]
\textbf{Koopmans'}\\
C$'$ & - & - & - & - & 0.0	& -0.1	& -0.1 & - &	-0.2	& -0.1 &	0.0  & - \\
C$^\alpha$ 	& - & - & - & 3.5	& -1.5	& -1.6& 	-1.8 & - & 	-1.7 &	-1.6	 & -1.3 & -\\
C$^\beta$ 	& - & - & -3.2	& -3.1	& -2.6	& -2.4	& -2.7	& - & -2.8	& -2.8	& -2.8 & -\\
C$_2$ &	-2.6& 	-2.2	& -2.3	& -2.4	& -2.1	& -1.8	& -2.1	& - & -2.6	& -2.5 & -  & -\\
C$_3$ & -3.5& 	-3.3	& -3.1	& -3.2	& -2.9	& -2.6	& -3.0	& - & -3.1	& -3.0 & - & -\\
C$_{3\mathrm{a}}$ &	-3.5& 	-3.1	& -3.2	& -3.2	& -2.8	& -2.7 &	-3.1	& - & -3.4	& -3.2 & - & - \\
C$_4$ & - & -3.2	& -3.2	& -3.2	& -2.8	& -2.8	& -3.2 & - & - & - & - & - 	\\
C$_5$ &	- & -3.3	& -3.3	& -3.3	& -3.0	& -3.0	& -3.3 & - & - & - & -  & -	\\
C$_6$ &	- & -3.2	& -3.2	& -3.2	& -2.9	& -2.8	& -3.2 & - & - & -& - & - \\
C$_7$ & - & -3.0	 &-3.1	& -3.1	& -2.7	& -2.7	& -3.0	& -& -2.8 & - & -  & - \\
C$_{7\mathrm{a}}$ &	-2.6 & 	-2.2 & 	-2.3	& -2.3	& -1.9	& -1.8 & 	-2.2 & - & 	-2.2 & 	-2.4  & - & -\\
\cline{1-1} \cline{2-2} \cline{3-3} \cline{4-4} \cline{5-5}  \cline{6-9} \cline{10-10} \cline{11-11} \cline{12-13} \\[-2.5ex]
\textbf{$\Delta$SCF}\\
C$'$ & - & - & - & - & -0.3	& -0.3	& -0.3	& 0.0	& -0.4	& -0.3 &	0.0 & 0.0\\
C$^\alpha$ 	& - & - & - & -3.9	& -2.4	& -2.5	& -2.7	& -1.3	& -2.5	& -2.4	& -1.8 & -1.7\\
C$^\beta$ 	& - & - & -3.6	& -3.7	& -3.4	& -3.3	& -3.5	& -2.3 &	-3.5	& -3.5	& -3.0 & -3.1\\
C$_2$ &	-3.5 & 	-3.4 &	-3.7 &	-3.8 &	-3.5 &	-3.3 &	-3.6 &	-2.7 &	-3.8 &	-3.6 & -  & -\\
C$_3$ & -4.3	& -4.5	& -4.4	& -4.5	& -4.2	& -4.0	& -4.3	& -3.3	& -4.3	& -4.1 & -  & -\\
C$_{3\mathrm{a}}$ &	-4.3 &	-4.2 &	-4.4	& -4.4	& -4.1	& -4.0 &	-4.3 &	-3.2 &	-4.5	& -4.2 & - & - \\
C$_4$ & - & -4.4& 	-4.5	& -4.6	& -4.2	& -4.2	& -4.5	& -3.4 & -  &  - & - & - \\
C$_5$ &	- & -4.5	& -4.6	& -4.6	& -4.3	& -4.3	& -4.6	& -3.4 &  - & - & - 	\\
C$_6$ &	- & -4.5& 	-4.6	& -4.6	& -4.3	& -4.2	& -4.5	& -3.3 & - & - & - \\
C$_7$ & - & -4.2	& -4.3	& -4.3	& -4.1	& -4.0	& -4.3	& -3.2 & 	-3.1 & - & - \\
C$_{7\mathrm{a}}$ &	-3.5 &	-3.4 &	-3.5 &	-3.5 &	-3.2 &	-3.1 &	-3.4 &	-2.3 &	-3.4 &	-3.4 & - \\
\hline\hline
\end{tabular*}
\end{threeparttable}
\end{table*}

\begin{table*}[h]
\centering
\begin{threeparttable}
\caption{\label{tab:c_his_series_bes} C~1\textit{s} PBE-calculated BEs for the His series of gas phase amino acid conformers, the extracted His molecule, and the subspecies molecules, as well as the solid state calculations. Gas phase BEs are relative to Ala C$'$, while solid state BEs are relative to C$'$ of that amino acid.}
\begin{tabular*} {1.0\textwidth}{l @{\extracolsep{\fill}} rrrrrrrrr}
\hline\hline

 &	\textbf{31}	& \textbf{32} & 	\textbf{33} &  \multicolumn{4}{c}{\textbf{His}} &   \textbf{Ala}\\
& conf 1	& conf 1	& conf 1 	& conf 1	& conf 2	& extract	& solid 	 & conf 1 & solid\\
\cline{1-1} \cline{2-2} \cline{3-3} \cline{4-4}  \cline{5-8} \cline{9-10}  \\[-2.5ex]
\textbf{Koopmans'}\\
C$'$ & - & - & - & -0.2& 	-0.2	& -0.3	& - & 0.0 & - \\
C$^\alpha$ 	& - & - & -3.6	& -1.6	& -1.7	& -2.1	& - & -1.3 & - \\
C$^\beta$ 	& - & -3.4	& -3.2	& -2.7	& -2.5	& -2.7	& - & -2.8 & - \\
C$_2$ &	-1.7	& -1.8	& -1.8	& -1.3	& -1.2	& -1.3 & - & - & -\\
C$_4$ & -2.7 & 	-2.6	& -2.7	& -2.2	& -2.0	& -2.1 & - & - 	& -\\
C$_5$ &	-2.2	& -2.4	& -2.4	& -2.1	& -1.8	& -2.0 & -  & - & -\\
\cline{1-1} \cline{2-2} \cline{3-3} \cline{4-4}  \cline{5-8} \cline{9-10}  \\[-2.5ex]
\textbf{$\Delta$SCF}\\
C$'$ & - & - & - & -0.4	& -0.5	& -0.6	& 0.0	& 0.0 & 0.0\\
C$^\alpha$ 	& - & - & -4.0 &	-2.4	& -2.5	& -2.8	& -1.8	& -1.8 & -1.7\\
C$^\beta$ 	& - & -3.6	& -3.7	& -3.3	& -3.2	& -3.3	& -2.7	& -3.0 & -3.1\\
C$_2$ &	-2.3	& -2.6	& -2.6	& -2.3	& -2.1	& -2.2	& -2.1 & - & - \\
C$_4$ & -3.4	& -3.4	& -3.6	& -3.2	& -3.0	& -3.1	& -2.8 & - & - \\
C$_5$ &	-2.9	& -3.3	& -3.4	& -3.1	& -2.9	& -3.0	& -2.8 & - & - \\
\hline\hline
\end{tabular*}
\end{threeparttable}
\end{table*}

\begin{table*}[h]
\centering
\begin{threeparttable}
\caption{\label{tab:c_extra_series_bes} C~1\textit{s} PBE-calculated BEs for the series of additional gas phase molecules, and conformer and solid state calculations for Ala. BEs are relative to Ala C$'$ in gas phase (solid state) for gas phase conformer (Ala solid state) calculations.}
\begin{tabular*} {1.0\textwidth}{l @{\extracolsep{\fill}} rrrrrrrrrrr}
\hline\hline

 &	\textbf{43}	& \textbf{42} & 	\textbf{41} &     \multicolumn{2}{c}{\textbf{Ala}} & 
 \textbf{44} & 	\textbf{2} &
 \textbf{45} & 	\textbf{46} &
 \textbf{47} & 	\textbf{11} \\
& conf 1	& conf 1	& conf 1 	& conf 1 & solid	& conf 1& conf 1& conf 1 & conf 1	& conf 1	& conf 1 \\
\cline{1-1} \cline{2-2} \cline{3-3} \cline{4-4}  \cline{5-6} \cline{7-7} \cline{8-8} \cline{9-9} \cline{10-10} \cline{11-11}  \cline{12-12} \\[-2.5ex]
\textbf{Koopmans'}\\
C$'$ & - & - & - & 0.0	& - & - & - & -	& -	& - & - \\
C$^\alpha$ 	& -2.6	& -2.6	& -2.5	& -1.3	& - & - & - & -	& -	& - & - \\
C$^\beta$ 	& -3.5 &	-3.4	& -3.5	& -2.8	& - & -3.5 &	-3.1 & - & - & - & - \\
C$_1$ &	- & - & - & - & - & -3.1 &	-2.7	& -2.5	& -1.9	& -3.2 &	-3.1\\
C$_2$ & - & - & - & - & - & -3.4	& -3.1	& -3.4	& -3.3	& -3.2 &	-2.9	\\
C$_3$ & - & - & - & - & - &  -3.3	& -3.0	& -3.3	& -3.2	& -3.3 &	-3.0\\
C$_4$ & - & - & - & - & - & -3.4 &	-3.1	& -3.3	& -3.5	& -1.9	 & -1.3\\
C$_5$ &	- & - & - & - & - & -3.3 &	-3.0	& -3.3	& -3.2	& -3.2 &	-2.9\\
C$_6$ & - & - & - & - & - & -3.4 &	-3.1	& -3.4	& -3.3	& -3.1 &	-2.8	\\
\cline{1-1} \cline{2-2} \cline{3-3} \cline{4-4}  \cline{5-6} \cline{7-7} \cline{8-8} \cline{9-9} \cline{10-10} \cline{11-11}  \cline{12-12} \\[-2.5ex]
\textbf{$\Delta$SCF}\\
C$'$ &  - & - & - & 0.0	& 0.0 & - & - & 	-	& -	& - & 	-\\
C$^\alpha$ 	& -3.2	& -3.0	& -2.7	& -1.8	& -1.7 & - & - &	- & 	- &	- & 	-\\
C$^\beta$ 	& -3.8 &	-3.6	& -3.6	& -3.0	& -3.1 & -3.9 & -3.5 & - & - & - & - \\
C$_1$ & - & - & - & - & - &	-3.9 &	-3.8 &	-3.1 &	-2.9 &	-3.7	 & -4.2\\
C$_2$ &	- & - & - & - & - & -4.0 &	-4.2	& -4.0	& -4.4	& -3.7 &	-3.9\\
C$_3$ &- & - & - & - & - &	-3.9 &	-4.1	& -3.8	& -4.2	& -3.9 &	-4.1\\
C$_4$ &- & - & - & - & - & -3.9	& -4.1	& -3.9	& -4.6	& -2.4	& -2.2\\
C$_5$ &	- & - & - & - & - & -3.9 &	-4.1	& -3.8	& -4.2	& -3.8	& -3.9\\
C$_6$ &	- & - & - & - & - &	-4.0 &	-4.2	& -4.0	& -4.4	& -3.7 &	-3.9\\
\hline\hline
\end{tabular*}
\end{threeparttable}
\end{table*}

\clearpage

\begin{table*}[h]
\centering
\begin{threeparttable}
\caption{\label{tab:n_trp_series_bes} N~1\textit{s} PBE-calculated BEs for the Trp series of gas phase amino acid conformers, the extracted Trp molecule, and the subspecies molecules, as well as the solid state calculations. Gas phase BEs are relative to Ala C$'$, while solid state BEs are relative to C$'$ of that amino acid.}
\begin{tabular*} {1.0\textwidth}{l @{\extracolsep{\fill}} rrrrrrrrrrrr}
\hline\hline

 &	\textbf{21}	& \textbf{22} & 	\textbf{23} & 	\textbf{24} & \multicolumn{4}{c}{\textbf{Trp}} &  	\textbf{25} & 	\textbf{26}& \multicolumn{2}{c}{\textbf{Ala}}\\
& conf 1	& conf 1	& conf 1 	& conf 1	& conf 1	& conf 2	& extract	& solid 	& conf 1 	& conf 1 & conf 1 & solid\\
\cline{1-1} \cline{2-2} \cline{3-3} \cline{4-4} \cline{5-5}  \cline{6-9} \cline{10-10} \cline{11-11} \cline{12-13} \\[-2.5ex]
\textbf{Koopmans'}\\
N$_1$ & - & - & - & - &  -0.3	& -0.3	& -0.2	& - & -0.4	& -0.3 & 	0.0 & -\\
N$_2$ & 0.5	& 0.6	& 0.5	& 0.5	&0.8	& 1.0	& 0.9& - & 	0.6 &	0.8 & - & - \\
\cline{1-1} \cline{2-2} \cline{3-3} \cline{4-4} \cline{5-5}  \cline{6-9} \cline{10-10} \cline{11-11} \cline{12-13} \\[-2.5ex]
\textbf{$\Delta$SCF}\\
N$_1$ &	 - & - & - & - & -0.6	& -0.6	& -0.4	& 0.0	& -0.7 &	-0.5	& 0.0 & 0.0\\
N$_2$ & 0.2	& -0.1	& -0.3	& -0.3	& -0.1	& 0.1	& 0.1	& -1.2	& 0.0	& 0.2 & - & - \\
\hline\hline
\end{tabular*}
\end{threeparttable}
\end{table*}

\begin{table*}[h]
\centering
\begin{threeparttable}
\caption{\label{tab:n_his_series_bes} N~1\textit{s} PBE-calculated BEs for the His series of gas phase amino acid conformers, the extracted His molecule, and the subspecies molecules, as well as the solid state calculations. Gas phase BEs are relative to Ala C$'$, while solid state BEs are relative to C$'$ of that amino acid.}
\begin{tabular*} {1.0\textwidth}{l @{\extracolsep{\fill}} rrrrrrrrr}
\hline\hline

 &	\textbf{31}	& \textbf{32} & 	\textbf{33} &  \multicolumn{4}{c}{\textbf{His}} &   \textbf{Ala}\\
& conf 1	& conf 1	& conf 1 	& conf 1	& conf 2	& extract	& solid 	 & conf 1 & solid\\
\cline{1-1} \cline{2-2} \cline{3-3} \cline{4-4}  \cline{5-8} \cline{9-10}  \\[-2.5ex]
\textbf{Koopmans'}\\
N$_1$ & - & - & - & -0.5	& -0.6	& -1.0	& - & 0.0 & - \\
N$_2$ & 1.0	& 0.8	& 0.9	& 1.2	& 1.4	& 1.6 & - & - & - \\
N$_3$ & -1.2	& -1.4	& -1.3	& -0.8	& -0.6	& -0.4 & - & - & - \\
\cline{1-1} \cline{2-2} \cline{3-3} \cline{4-4}  \cline{5-8} \cline{9-10}  \\[-2.5ex]
\textbf{$\Delta$SCF}\\
N$_1$ & - & - & - & -0.8 & 	-0.8	& -1.3	& 0.0	& 0.0 & 0.0\\
N$_2$ & 0.8 & 	0.5	& 0.4	& 0.8	& 1.0	& 1.2	& -0.6 & - & - \\
N$_3$ & -1.6 &	-1.9	& -2.0	& -1.5	& -1.4	& -1.1	& -2.2 & - & - \\
\hline\hline
\end{tabular*}
\end{threeparttable}
\end{table*}

\begin{table*}[h]
\centering
\begin{threeparttable}
\caption{\label{tab:n_extra_series_bes} N~1\textit{s} BEs for the series of additional gas phase molecules, and conformer and solid state calculations for Ala. BEs are relative to Ala N$^1$ in gas phase (solid state) for gas phase conformer (Ala solid state) calculations.}
\begin{tabular*} {1.0\textwidth}{l @{\extracolsep{\fill}} rrrrrrrrrrr}
\hline\hline

 &	\textbf{43}	& \textbf{42} & 	\textbf{41} &     \multicolumn{2}{c}{\textbf{Ala}} & 
 \textbf{44} & 	\textbf{2} &
 \textbf{45} & 	\textbf{46} &
 \textbf{47} & 	\textbf{11} \\
& conf 1	& conf 1	& conf 1 	& conf 1 & solid	& conf 1& conf 1& conf 1 & conf 1	& conf 1	& conf 1 \\
\cline{1-1} \cline{2-2} \cline{3-3} \cline{4-4}  \cline{5-6} \cline{7-7} \cline{8-8} \cline{9-9} \cline{10-10} \cline{11-11} \cline{12-12}  \\[-2.5ex]
\textbf{Koopmans'}\\
N$_1$ & - & - & -1.3 &	0.0	& - & - & - & -1.3	& -0.4& - & -  \\
N$_2$ & - & -1.2 & - & - & - & - & - & -& -  & -  & -  \\
N$_3$ & -1.0 & - & - & - & - & - & - & -& - & -& -  \\
\cline{1-1} \cline{2-2} \cline{3-3} \cline{4-4}  \cline{5-6} \cline{7-7} \cline{8-8} \cline{9-9} \cline{10-10} \cline{11-11} \cline{12-12}  \\[-2.5ex]
\textbf{$\Delta$SCF}\\
N$_1$ & - & - & -1.1 & 0.0 &	0.0 &	- & - &  -1.4	& -0.7 & - & - \\
N$_2$ & - & -1.5	& - & - & - & - & - & - & - & - & - \\
N$_3$ &	-1.7 & - & - & - & - & - & - & - & - & - & -\\
\hline\hline
\end{tabular*}
\end{threeparttable}
\end{table*}

\clearpage

\begin{table*}[h]
\centering
\begin{threeparttable}
\caption{\label{tab:exp_bes} Absolute experimental BEs for solid state amino acids as extracted from peak fit analysis. The Ala results have been previously reported\cite{Pi2020}.}
\begin{tabular*} {1.0\textwidth}{l @{\extracolsep{\fill}} rrrrr}

\hline\hline

 & Ala & Phe & Tyr & Trp & His \\
 
\cline{1-1} \cline{2-2} \cline{3-3} \cline{4-4} \cline{5-5} \cline{6-6} \\[-2.5ex]

C$'$              & 288.4 & 288.5 & 288.7 & 288.8 & 288.4 \\
C$^\alpha$        & 286.5 & 286.6 & 286.2 & 286.7 & 286.5 \\
C$^\beta$         & 285.2 & 285.9 & 285.1 & 285.8 & 285.2 \\
C$_1$             & -     & 285.4 & 284.4 & -     & - \\
C$_2$             & -     & 284.8 &	284.4 &	285.2 &	286.0 \\
C$_3$             & -     & 284.8 & 284.4 & 284.5 & - \\
C$_{3\mathrm{a}}$ & -     & -     & -     & 284.5 & - \\
C$_4$             & -     &	284.8 & 285.6 & 284.5 & 285.2 \\
C$_5$             & -     & 284.8 &	284.4 & 284.5 & 285.2 \\
C$_6$             & -     & 284.8 & 284.4 & 284.5 & - \\
C$_7$             & -     & -     & -     & 284.5 & -\\
C$_{7\mathrm{a}}$ & -     & -     & -     & 285.8 & -\\

\cline{1-1} \cline{2-2} \cline{3-3} \cline{4-4} \cline{5-5} \cline{6-6} \\[-2.5ex]

O$^1$ & 531.4 & 531.4 & 531.7 & 531.5 & 531.3 \\
O$^2$ & 531.4 & 531.4 & 531.7 & 531.5 & 531.3 \\
O$^3$ & - & - & 532.9 & - & - \\

\cline{1-1} \cline{2-2} \cline{3-3} \cline{4-4} \cline{5-5} \cline{6-6} \\[-2.5ex]

N$^1$ & 401.4 & 401.3 & 401.6 & 401.6 & 401.5 \\
N$^2$ & -     & -     & -     & 400.2 & 400.4 \\
N$^3$ & -     & -     & -     & -     & 398.8 \\

\hline\hline
\end{tabular*}

\end{threeparttable}
\end{table*}

\begin{figure*}[ht]
\centering
\includegraphics[width=1.0\textwidth]{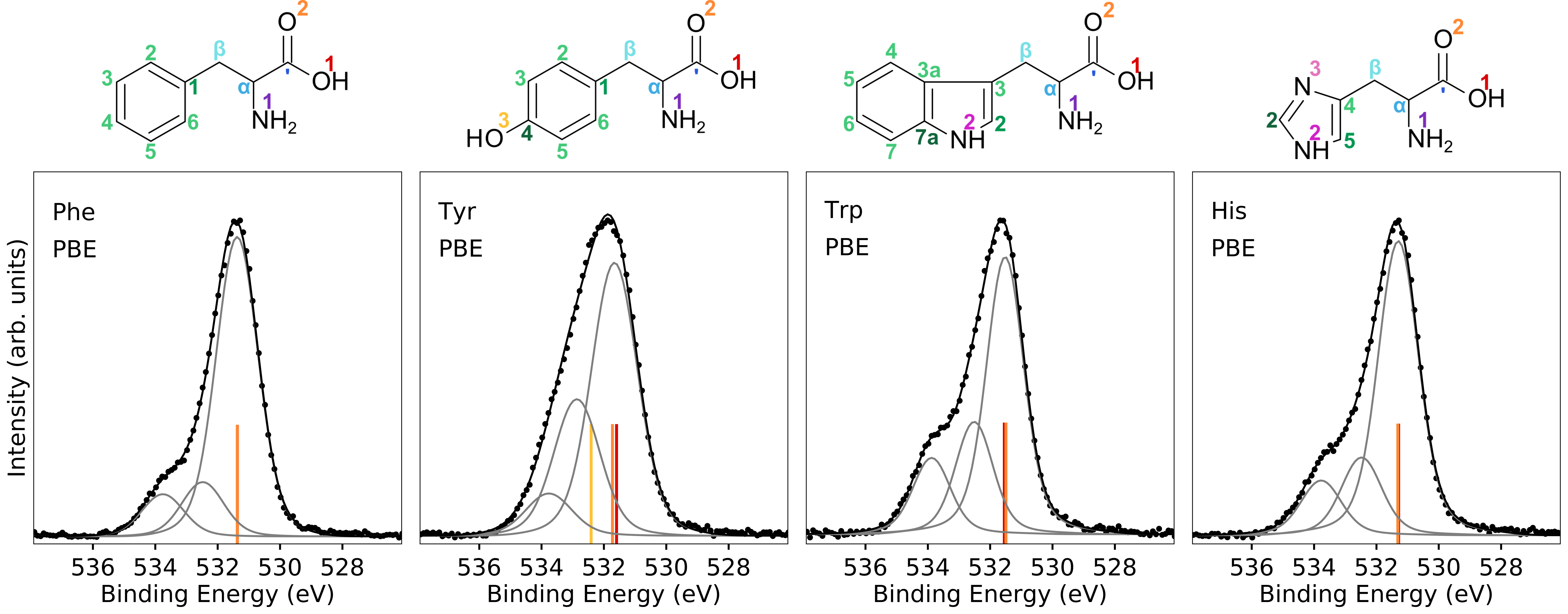}
\caption{O~1\textit{s} core level spectra with experiments depicted as black dots, experimental peak fits denoted as grey/black solid lines, and calculated BEs shown as coloured vertical lines. PBE0 calculations are omitted due to the similarity with PBE results. Calculated BEs have been aligned with the experimental spectra by aligning with respect to the lowest BE peak, taking the average calculated BE of O$^1$ and O$^2$. \label{fig:O_spectra}}
\end{figure*}

\vspace{-5pt}
\bibliographystyle{iopart-num}
\bibliography{references}